\documentclass{aastex62}

\usepackage{amsmath}
\usepackage{array}
\shorttitle{NEI on EUV and X-ray observations}
\shortauthors{Lee et al.}

\begin{document}

\title{Non-equilibrium ionization effects on solar EUV and X-ray imaging observations
}

\email{jlee@khu.ac.kr}

\author{Jin-Yi Lee}
\affil{Department of Astronomy and Space Science, Kyung Hee University, 
Yongin-si, Gyeonggi-do, 17104, Republic of Korea }

\author{John C. Raymond}
\affiliation{Harvard-Smithsonian Center for Astrophysics, 
Cambridge, MA 02138, USA}

\author{Katharine K. Reeves}
\affiliation{Harvard-Smithsonian Center for Astrophysics, 
Cambridge, MA 02138, USA}

\author{Chengcai Shen}
\affiliation{Harvard-Smithsonian Center for Astrophysics, 
Cambridge, MA 02138, USA}

\author{Yong-Jae Moon}
\affil{Department of Astronomy and Space Science, Kyung Hee University, 
Yongin-si, Gyeonggi-do, 17104, Republic of Korea }
\affiliation{School of Space Research, Kyung Hee University, 
Yongin-si, Gyeonggi-do, 17104, Republic of Korea}

\author{Yeon-Han Kim}
\affiliation{Korea Astronomy \& Space Science Institute, 
Daejeon, 34055, Republic of Korea}
\affiliation{University of Science and Technology, Daejeon, 34113, Republic of Korea}

\begin{abstract}

During transient events such as major solar eruptions, the plasma can be far from the equilibrium ionization state because of rapid heating or cooling. Non-equilibrium ionization~(NEI) is important in rapidly evolving systems where the thermodynamical timescale is shorter than the ionization or recombination time scales. We investigate the effects of NEI on EUV and X-ray observations by the Atmospheric Imaging Assembly (AIA) on board Solar Dynamic Observatory and X-ray Telescope (XRT) on board Hinode. Our model assumes that the plasma is initially in ionization equilibrium at low temperature, and it is heated rapidly by a shock or magnetic reconnection.  We tabulate the responses of the AIA and XRT passbands as functions of temperature and a characteristic timescale, $n_{e}t$. We find that most of the ions reach equilibrium at $n_{e}t\leq$10$^{12}$ cm$^{-3}$s. Comparing ratios of the responses between different passbands allows us 
to determine whether a combination of plasmas at temperatures in ionization equilibrium can account for a given AIA and XRT observation. It also expresses how far the observed plasma is from equilibrium ionization. We apply the ratios to a supra-arcade plasma sheet on 2012 January 27. We find that the closer the plasma is to the arcade, the closer it is to a single-temperature plasma in ionization equilibrium. 
We also utilize the set of responses to estimate the temperature and density for shocked plasma associated with a coronal mass ejection on 2010 June 13.  The temperature and density ranges we obtain are in reasonable agreement with previous works. 
\end{abstract}

\keywords{Sun: activity --- Sun: corona --- Sun: coronal mass ejections (CMEs) 
--- Sun: UV radiation --- Sun: X-rays, gamma rays --- Plasmas --- Shock waves}

\section{Introduction} \label{sec:intro}

Study of the physical properties of erupting solar coronal plasma is needed to understand the mechanisms of solar eruptions. 
Recent high temporal and spatial resolution measurements make the detailed analysis of erupting solar coronal plasma possible. 
Most coronal analyses assume ionization equilibrium to determine the physical properties of the erupting plasma 
\citep{cheng2012, hannah2013, patsourakos2013, tripathi2013, hanneman2014, lee2015, lee2017, reeves2017}. 
In ionization equilibrium, the responses of the Atmospheric Imaging Assembly (AIA) on board {\it Solar Dynamic Observatory} and 
X-ray Telescope (XRT) on board {\it Hinode} are functions of temperature alone. 
However, if the thermodynamical timescale in a rapidly evolving system is shorter than 
the ionization and recombination timescale, then the plasma is out of equilibrium ionization (EI) \citep[e.g.][]{cox1972, shapiro1977, golub1989, dudik2017}. 
In that case, the instrument responses are functions of temperature and a characteristic timescale, $n_{e}t$, 
which are the parameters of the time-dependent ionization equation \citep[e.g.,][]{shen2015}, where $n_e$ and $t$ are density and time, respectively.

A time-dependent ionization model \citep{shen2015} 
performs fast calculations 
by an eigenvalue method \citep{masai1984, hughes1985}. 
The model pre-computes the ionization and recombination rates, and 
those are saved into tables of eigenvectors and eigenvalues. 
Therefore, the model can calculate ion fractions 
easily for a large grid of temperature and $n_{e}t$. 
Models using the eigenvalue method have been used for rapidly heated astrophysical plasma  
such as supernova remnants \citep{hughes1985, kaastra1993}, current sheets \citep{shen2013}, and streamers \citep{shen2017}. 
\citet{smith2010} present the characteristic timescale, $n_{e}t$, to reach equilibrium for astrophysically abundant elements. 
This gives a rough idea whether the plasma with a combination of $n_e$ and $t$ is in equilibrium.  
The importance of non-equilibrium ionization~(NEI) effects in the solar atmosphere has been addressed in 
specific cases (see references in \citealp{bradshaw2013, dudik2017}).  

In this analysis, we investigate the effects of NEI on EUV and X-ray observations 
by AIA and XRT. For the investigation, first, we obtain the ion fractions for all the ions 
as a function of temperature and a characteristic timescale, $n_{e}t$, using a time-dependent ionization model \citep{shen2015}. 
Second, we calculate the emissivities for all the lines of ions of abundant elements using CHIANTI~8.07 \citep{delzanna2015}, 
and then we find the temperature response for each ion by multiplying the emissivities by the effective area for each AIA and XRT passband. 
Lastly, the ion fractions are multiplied by the temperature response for each passband, 
which results in a 2D grid for each combination of temperature and the characteristic time scale.
This set of passband responses is used for plasma that is rapidly ionized in a current sheet or a shock. 
We calculate the ratios between different passband responses to compare with the observations by AIA and XRT. 
We find that the ratio-ratio plots can be used to determine the departure of equilibrium as well as the constraints on temperature and density.  
As examples, we apply our results to a supra-arcade plasma sheet \citep{hanneman2014} and the shocked plasma in a 2010~June~13 CME \citep{ma2011, kozarev2011}.

In Section 2, we describe the calculations to obtain the set of passband responses for plasma in non-equilibrium. In Section 3, we show the ratios between passbands 
in NEI and the application to a supra-arcade plasma sheet and a shock event. In Section 4, we present our conclusions. 

\section{Calculations} \label{sec:calculation}

We find the temperature and characteristic responses of AIA and XRT using a time-dependent ionization model \citep{shen2015} and an atomic database CHIANTI 8.07 \citep{delzanna2015} in {\it SolarSoft} (SSW). 
This allows us to analyze the observations by the AIA and XRT in NEI states. 

\subsection{Ion fractions in NEI}
We obtain the ion fractions for all the ions of the 16 most abundant elements, H, He, C, N, O, Ne, Na, Mg, Al, Si, S, Ar, Ca, Cr, Fe, and Ni. 
Our model assumes that the plasma is in equilibrium at a low temperature in the initial state, 
and then it is rapidly heated in a short time. 
We calculate the ion fractions as a function of temperature and a characteristic timescale, 
$n_{e}t$, using a time-dependent ionization model \citep{shen2015}, which uses an eigenvalue method. 
The model pre-computes the ionization and recombination rates, and 
the rates are saved into tables containing the eigenvalues and eigenvectors for fast calculation. 
Using the pre-computed tables, the ion fractions are calculated with the time-dependent ionization equation, 

\begin{equation}
\frac {df_{i}}{dt} = n_{e} [C_{i-1} f_{i-1} - (C_{i} + R_{i})f_{i} + R_{i+1}f_{i+1}]
\end{equation}

\noindent where $f_i$ is ion fraction with charge state $i$, $n_e$ is density, and $t$ is time. $C_i$ and $R_i$ are ionization and recombination rate coefficients, 
which are taken from CHIANTI~8.07. 

\begin{figure}
\centering
\label{fig:time}
\includegraphics[width=80mm]{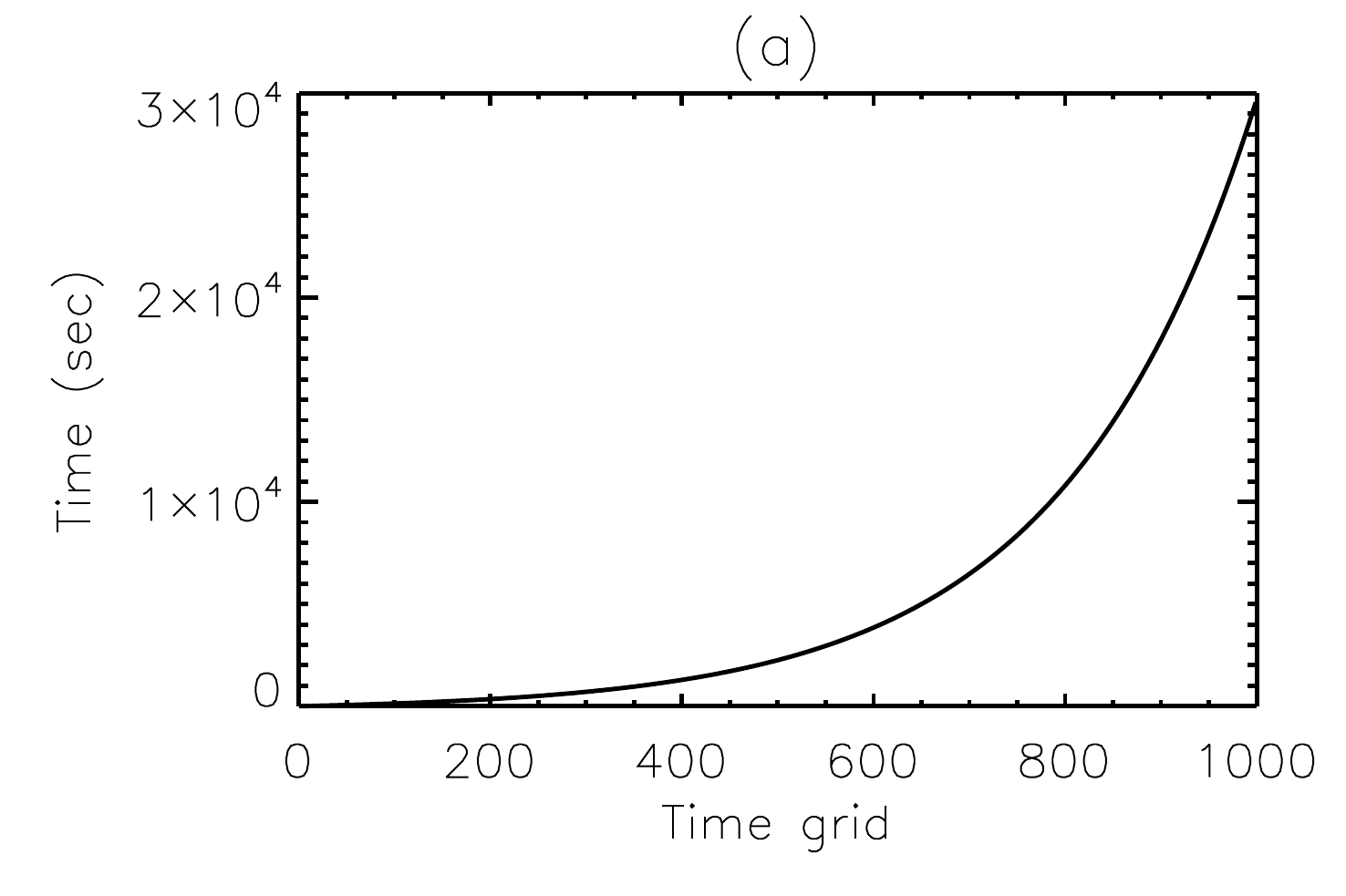}
\includegraphics[width=80mm]{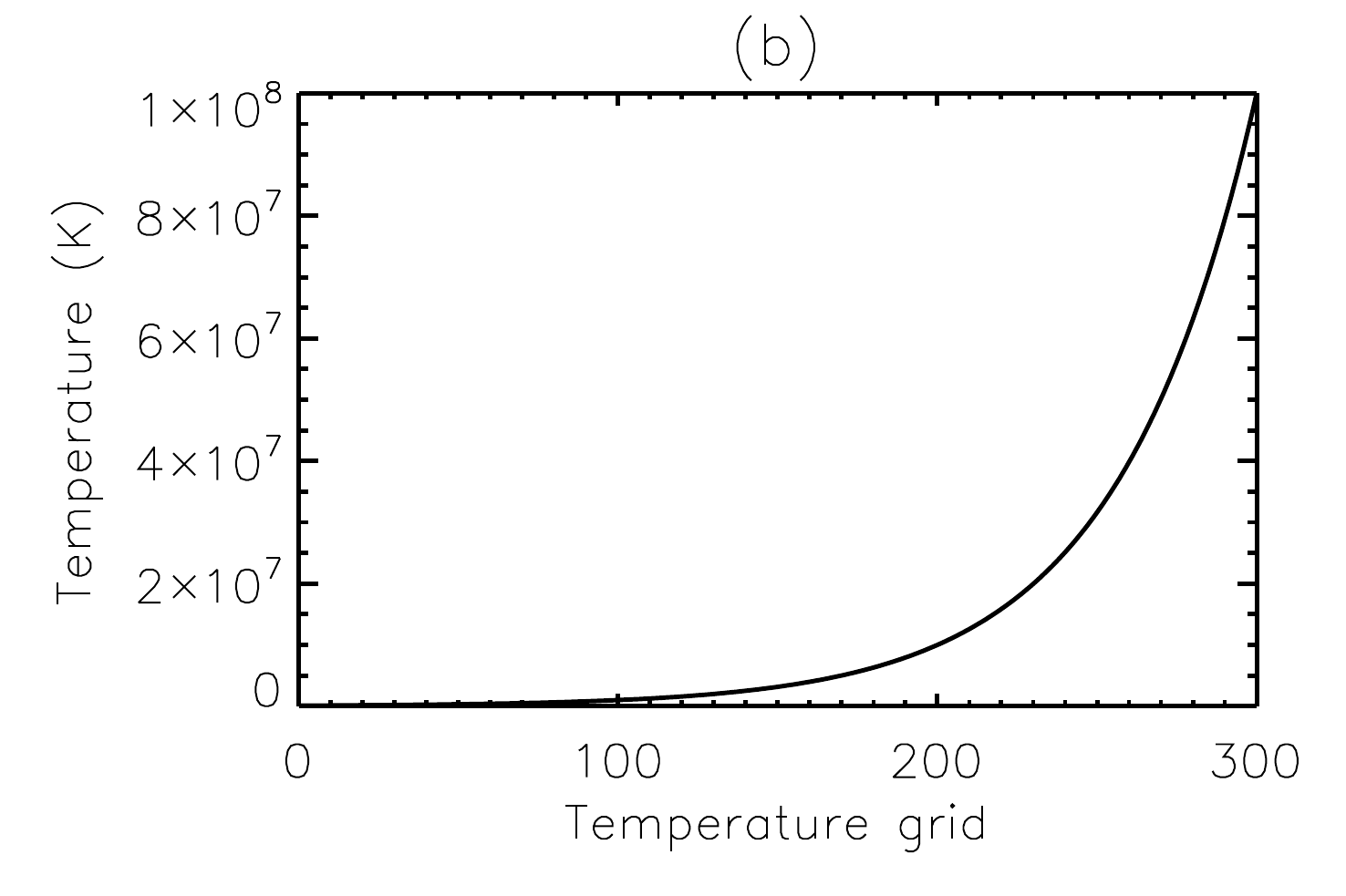}
\caption{Time and temperature grids} 
\end{figure}

The equation gives the ion fractions with a characteristic timescale, $n_{e}t$. 
We use time steps that increases exponentially to $\sim$3$\times$10$^4$~sec (Figure~\ref{fig:time}~(a)). 
We use a time step index of 1000, which increases exponentially with $t=t+dt$ to apply a larger set of time scales, where $dt$ is $e^{0.005\times{k}}$ and k is a time grid. 

Using a density, 1$\times$10$^8$ cm$^{-3}$, we calculate the ion fractions with the characteristic timescales from 1$\times$10$^8$ to 3$\times$10$^{12}$ cm$^{-3}$sec. 
The ionization and recombination rate coefficients are only very slightly density dependent at coronal densities \citep{vernazza1979}, and so the density dependence is ignored. 

The rate coefficients are functions of temperature. 
We assume an initial temperature of 10$^5$ K, and the plasma is heated to
10$^5$ to 10$^8$K at the first time step (Figure~\ref{fig:time}~(b)). 
The initial temperature is not important after a very short time provided that the temperature jump in substantial. 
We use a temperature step index of 300, which is the temperatures with 
10$^{(5 + (0.01)\times{k})}$ K, where k is a temperature grid. 
We find the ion fractions as a function of element (Z), ion (z), temperature (T), 
and charateristic timescale, $f(Z, z, T, n_{e}t)$. 

\begin{figure}
\centering
\includegraphics[width=80mm]{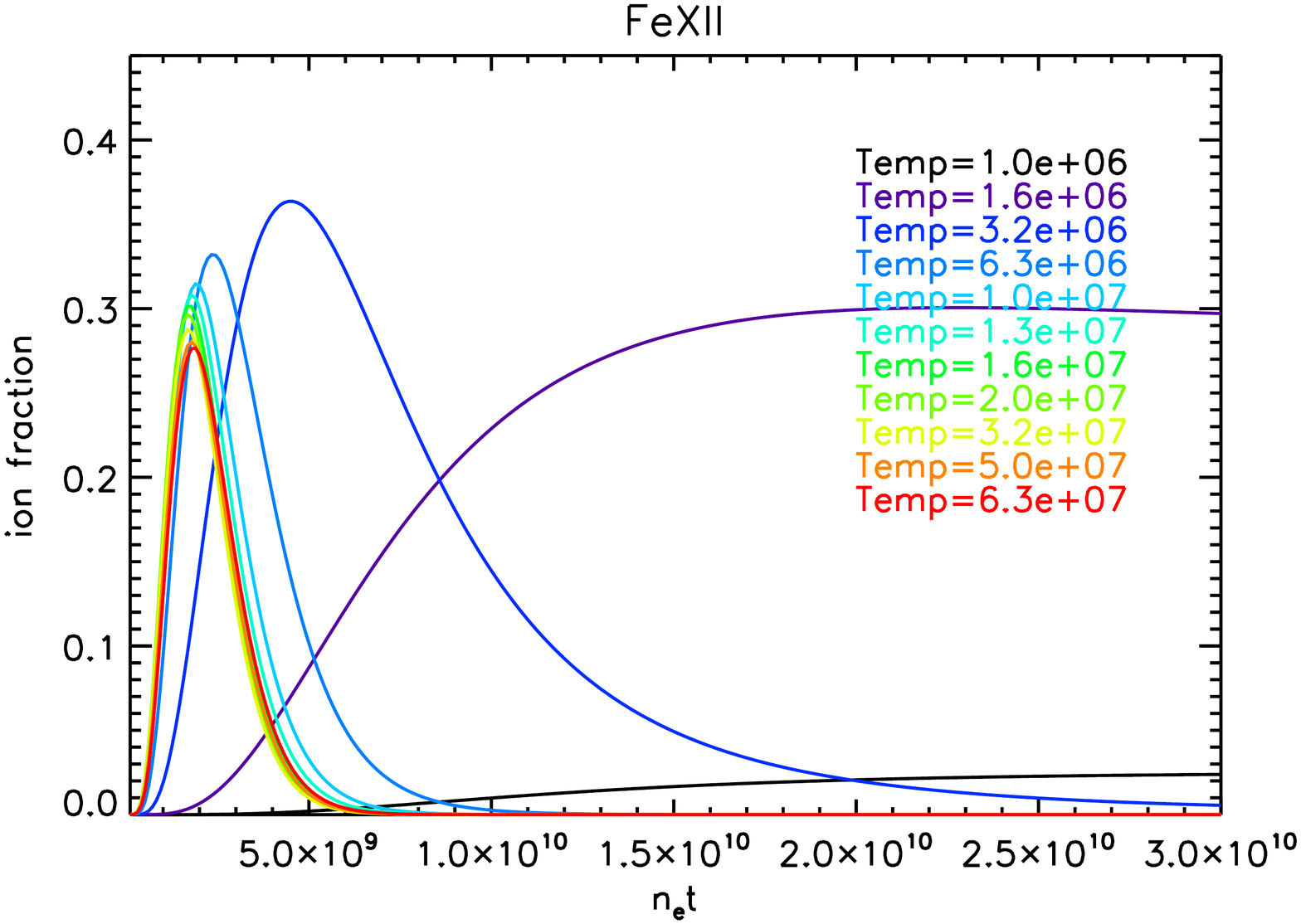}
\includegraphics[width=80mm]{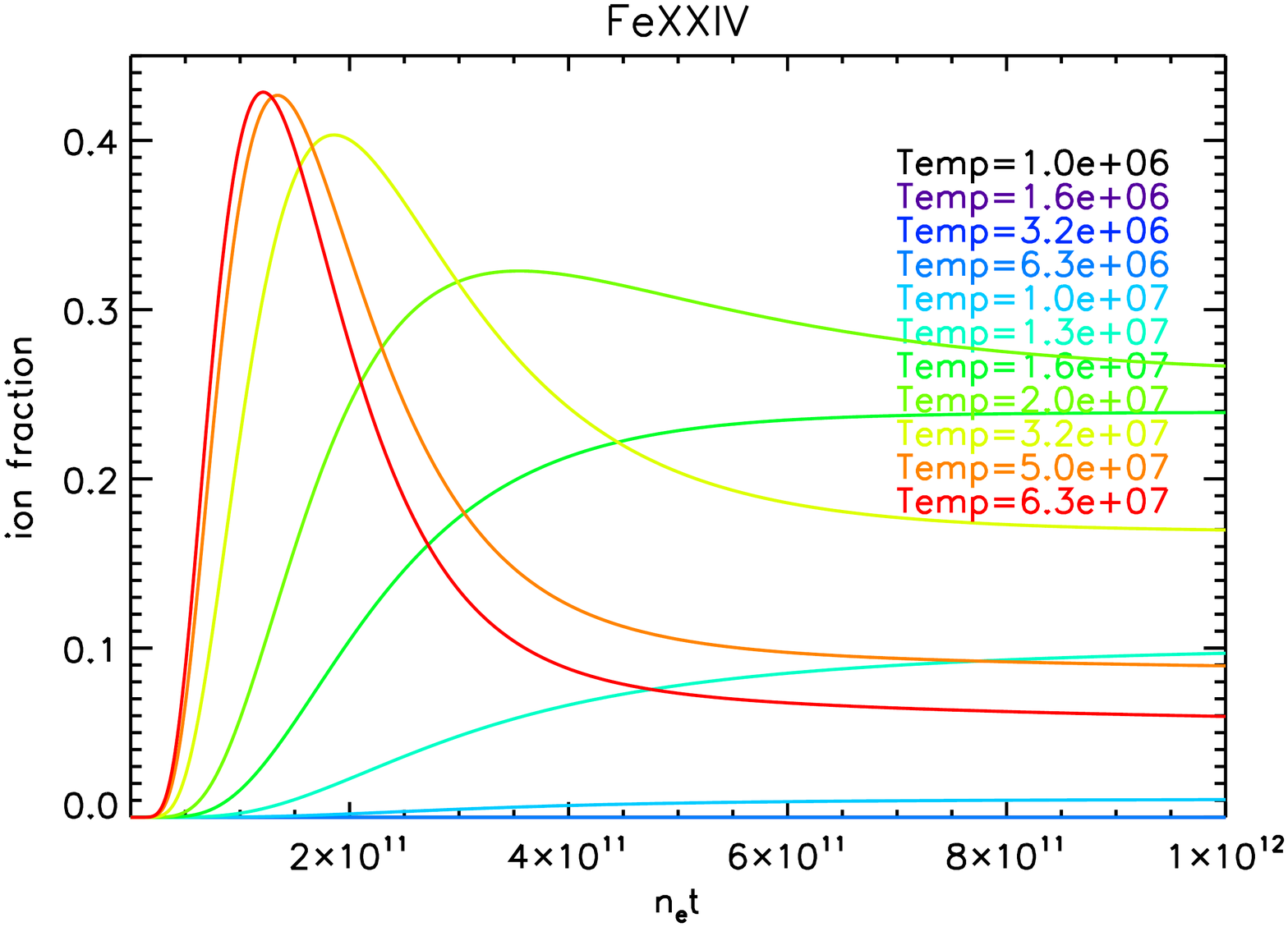}
\caption{$n_{e}t$ vs. Fe XII and Fe XXIV fractions. Colored lines represent different temperatures. 
These ions dominate the AIA~193 \AA\ passband except at very low T and $n_{e}t$. }
\label{fig:frac}
\end{figure}

Figure~\ref{fig:frac} shows the change of ion fractions with $n_{e}t$ with different temperatures from 1~MK to 63~MK for Fe~XII and Fe~XXIV, 
which are the dominant lines in the AIA 193 \AA\ passband \citep{lemen2012}. 
The Fe~XII fraction for the temperature of $\sim$1.6~MK (purple) evolves into equilibrium in 
$n_{e}t\leq$2$\times$10$^{10}$~cm$^{-3}$~sec while at higher temperatures, the fractions approach equilibrium earlier. 
The Fe~XXIV fraction for the temperature of $\sim$20~MK (green) evolves into equilibrium in 
$n_{e}t$=$\sim$8$\times$10$^{11}$~cm$^{-3}$sec. 
This indicates that the plasmas with 1.6~MK and 20~MK take about 2$\times$10$^2$~sec and 8$\times$10$^3$~sec to reach the equilibrium, respectively, for a plasma density of 10$^8$~cm$^{-3}$. 

\subsection{Emissivities}
 
We calculate the temperature responses for each element and ion so that 
they can be multiplied by the ion fractions,  
which are also functions of each element and ion. 
Firstly, we calculate the line emissivities for each ion of elements using atomic data from CHIANTI 8.07 \citep{delzanna2015}, 
and then we find the temperature responses by multiplying the line emissivities for each ion 
by the effective area for each AIA and XRT passband. 
The temperature response is given by, 

\begin{equation}
Resp(Z, z, T, band) = \sum_{i=1}^{nlines} \frac { \epsilon{(i, T)} } { 10^9 } A_{eff,band} (\lambda (i)), 
\end{equation}

\noindent where  $Resp(Z, z, T, band)$~[DN~cm$^5$~sec$^{-1}$] is the temperature response for each element ($Z$), ion ($z$), temperature ($T$), and the passband ($band$) of the AIA and XRT. 
We calculate the emissivities for the temperature range from 10$^5$~K to 10$^8$~K, 
which is the same range used in the calculation of ion fractions. 
We find the responses for the seven EUV passbands of 
AIA (94 \AA, 131 \AA, 171 \AA, 193 \AA, 211 \AA, 304 \AA, and 335 \AA) and 
all nine XRT passbands (Al\_mesh, Al\_poly, Al\_med, Al\_thick, Be\_thin, Be\_med, Be\_thick, C\_poly, and Ti\_poly). 
The line emissivity, $\epsilon(i, T)$ [photon~sec$^{-1}$], is calculated at an arbitrary density, $n_e$=10$^9~$cm$^{-3}$, by the procedure (emiss\_calc.pro) in CHIANTI~8.07. 
Then, the $\epsilon(i,T)$ is divided by the density. The choice of the density does not affect the calculation of the temperature responses 
because the density is an independent parameter for calculating the ion balance since we do not include photoexcitation and stimulated emission. 
We include the transitions by dielectronic recombination. 
The procedure calculates all the lines including the transitions 
where only theoretical energies are available for at least one of the two levels. 
Among these transitions, there are unrealistically high emissivities for a few lines at high temperatures in the CHIANTI data base. 
We exclude the emissivities of 58 lines (81 transitions) at higher temperature than 63~MK. 
{\it A$_{eff}$}~[cm$^2$ DN photon$^{-1}$] is the effective area for each passband ($band$) as a function of the wavelength ($\lambda$) for each transition line ($i$). The effective areas are calculated by procedures, aia\_get\_response.pro and make\_xrt\_wave\_resp.pro in SSW for AIA and XRT, respectively. The effective areas are calculated at a given specific date to consider the time-varing degradation of instruments \citep{boerner2014, narukage2011}. 
In this analysis, we calculate the responses on 2012 January 27 and also on 2010 June 13 for a shocked plasma in Section~3.2. 

\subsection{Temperature and characteristic timescale responses}

We find the temperature and characteristic timescale responses. This is the set of passband responses for plasma that is rapidly ionized in a current sheet or a shock.
The ion fractions are multiplied by the temperature responses calculated in Section~2.2, which results in a 2D grid for each combination of temperature and the characteristic timescale. 
The responses are given by.  

\begin{equation}
R(T, n_{e}t, band) = \sum_{Z} \sum_{z} Resp(Z,z,T,band) AB(Z) f(Z,z,T,n_{e}t) 
\end{equation}
 
\noindent where $R(T, n_{e}t, band)$ is the temperature (K) and characteristic timescale response (cm$^{-3}$sec), $AB(Z)$ is abundance, and $f(Z,z,T,n_e{t})$ is the ion fraction calculated using the time-dependent ionization model in Section~2.1. 
We use a coronal abundance (sun\_coronal\_1992\_feldman\_ext.abund) in CHIANTI \citep{feldman1992, landi2002, grevesse1998}.
Lastly, we add the continuum to the responses as below. 

\begin{equation}
R(T, n_{e}t, band) = R(T, n_{e}t, band) + \sum_{\lambda} cont(T, \lambda) A_{eff,band} (\lambda), 
\end{equation}

\noindent where $cont(T, \lambda)$ is the continuum calculated by procedures in CHIANTI~8.07, freefree.pro, freebound.pro, and two\_photon.pro for Bremsstrahlung emission,  free-bound emission, and, two-photon emission, respectively.  

Finally, we find the temperature and characteristic timescale responses in units of DN~cm$^5$sec$^{-1}$pix$^{-1}$ multiplying by $\Omega$/4$\pi$, 
where $\Omega$ is given by the pixel size, 0.6$''$ and 1.0286$''$ for AIA and XRT, respectively

\begin{figure}
\centering
\includegraphics[width=80mm]{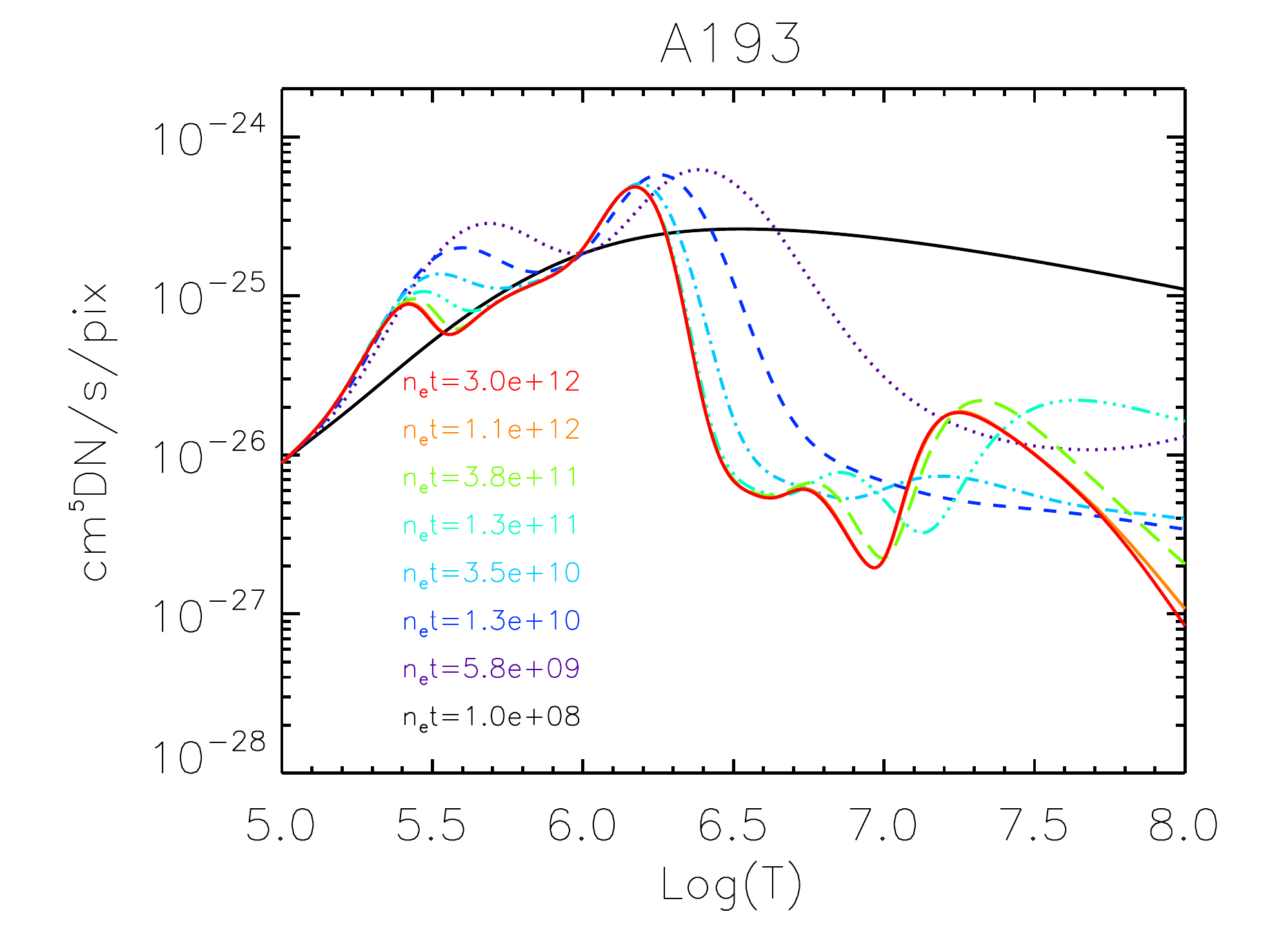}
\includegraphics[width=80mm]{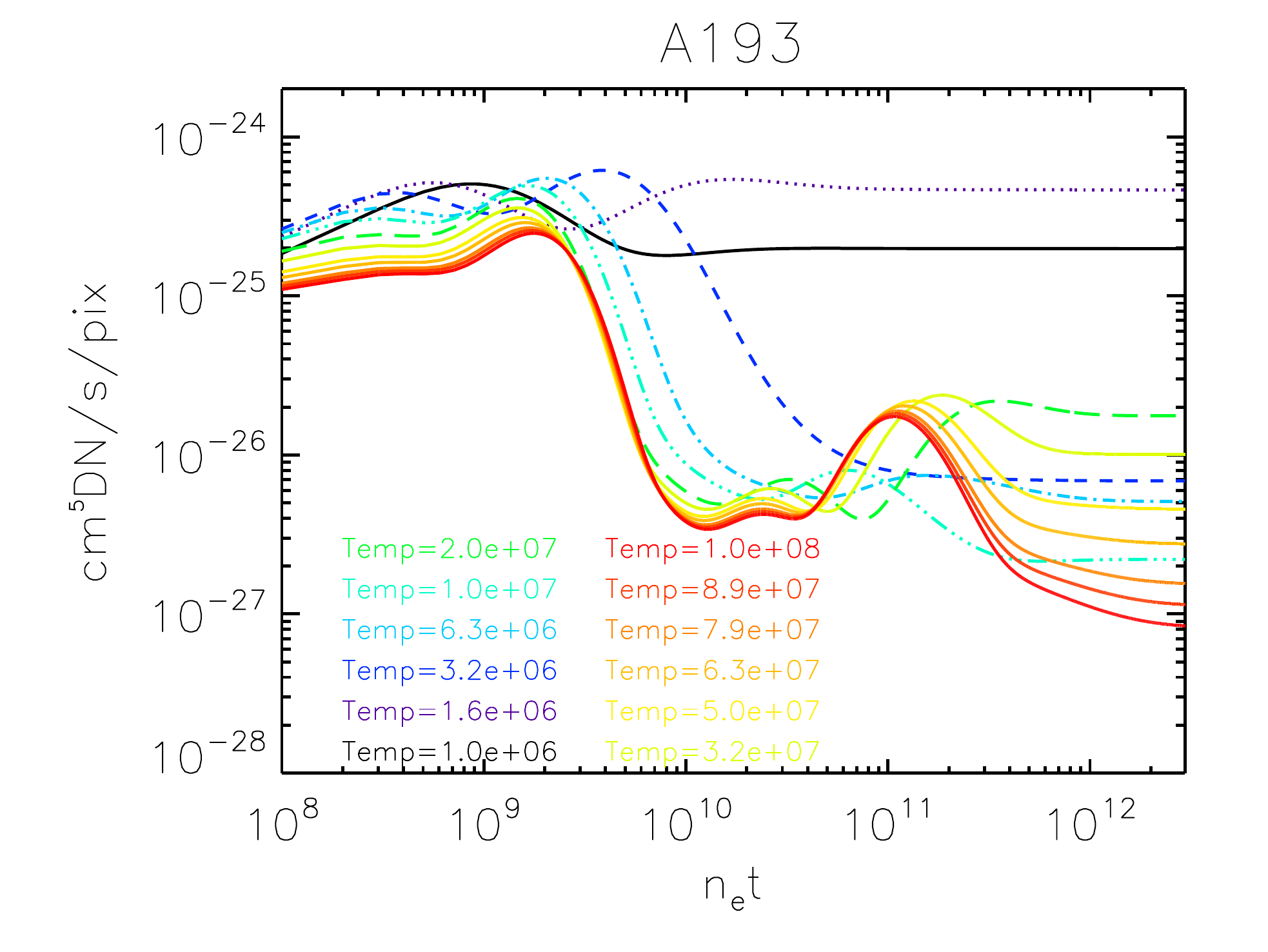}
\caption{Temperature and characteristic timescale responses for AIA 193 \AA. 
Colored lines represent characteristic timescales and temperatures in the left and right panels, respectively. 
Note that the compressed logarithmic scale of the Y-axis.}
\label{fig:aresp}
\end{figure}

\begin{figure}
\centering
\includegraphics[width=80mm]{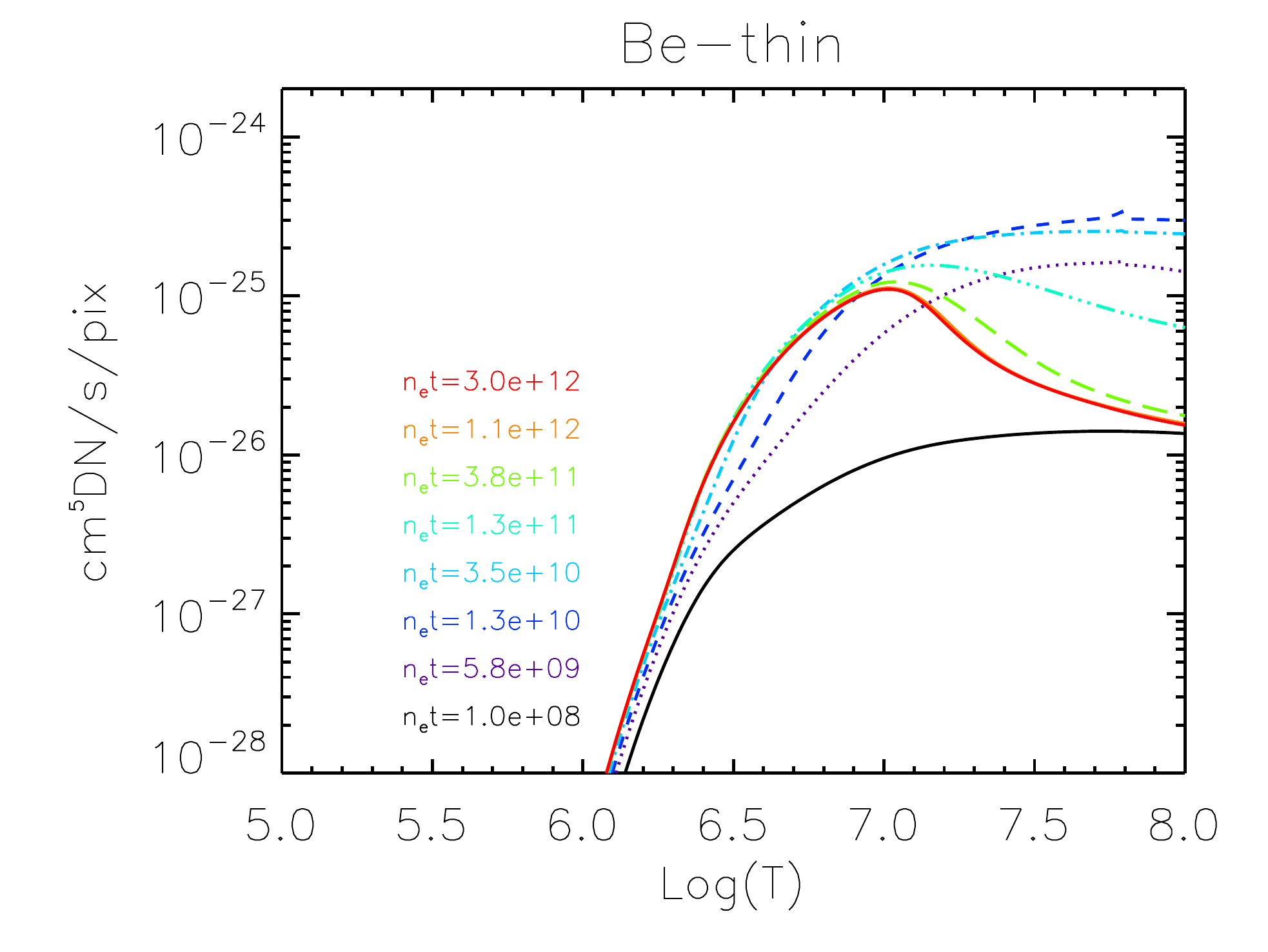}
\includegraphics[width=80mm]{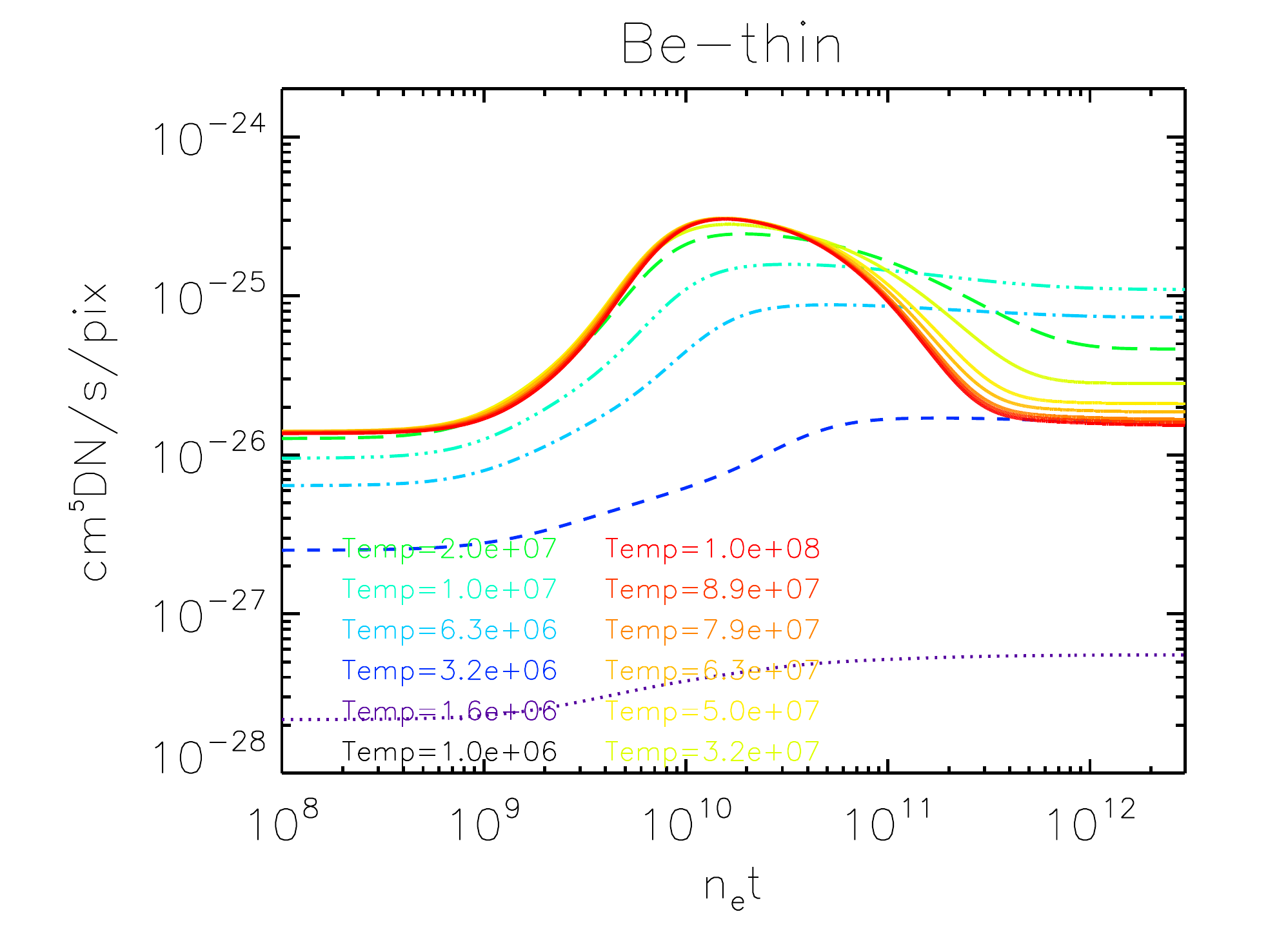}
\caption{Temperature and characteristic timescale responses for XRT Be$\_$thin. 
Colored lines represent characteristic timescales and temperatures in the left and right panels, respectively.}
\label{fig:xresp}
\end{figure}

As an example, we show the temperature and characteristic timescale responses 
for the AIA~193~\AA\ in Figure~\ref{fig:aresp}. 
In the left panel, the temperature response approaches the responses in equilibrium \citep{lemen2012, lee2017}  
with $n_{e}t\approx$10$^{12}$cm$^{-3}$sec. 
The red solid colors in the left panels of Figure~\ref{fig:aresp} and Figure~\ref{fig:xresp} represent the responses in equilibrium.  
The responses with small $n_{e}t$ (e.g. black and purple colors) at higher temperature  
are due to the contribution of O~V \citep{ciaravella2000, bryans2012, mccauley2013} and other low ionization species. 
In the right panel, a small bump at the larger $n_{e}t$ and higher temperature corresponds to the 
peak Fe~XXIV fraction around  $n_{e}t\approx$1$\times$10$^{11}$cm$^{-3}$sec in the right panel of Figure~\ref{fig:frac}. 
The temperature response of the XRT~Be\_thin also shows that 
the response goes to the response in equilibrium \citep{golub2007, lee2017} 
with around $n_{e}t\approx$10$^{12}$cm$^{-3}$sec (left panel in Figure~\ref{fig:xresp}). 
The responses show that most of ions are fully ionized at the highest temperature and large $n_{e}t$, 
and most of the high temperature emission is due to the Bremsstrahlung emission. 
In the right panel in Figure~\ref{fig:xresp}, the peak in Be\_thin at $n_{e}t$ = 10$^{10}$-10$^{11}$cm$^{-3}$sec 
is from emission lines that are strong in equilibrium at T=3$\sim$20~MK. 
 
\section{Results and Discussion} \label{sec:results}  
 
We show the ratios between different passbands to make available a comparison 
with the AIA and XRT observations and discuss the departure from equilibrium 
as well as the constraints on temperature and density. 
As examples to compare our results with observations, 
we apply the model to a supra-arcade plasma sheet~\citep{hanneman2014} and 
shocked plasma in the 2010~June~13 CME~\citep{ma2011, kozarev2011}. 

\subsection{Passband ratios in NEI}

\begin{figure}
\centering
\includegraphics[width=80mm]{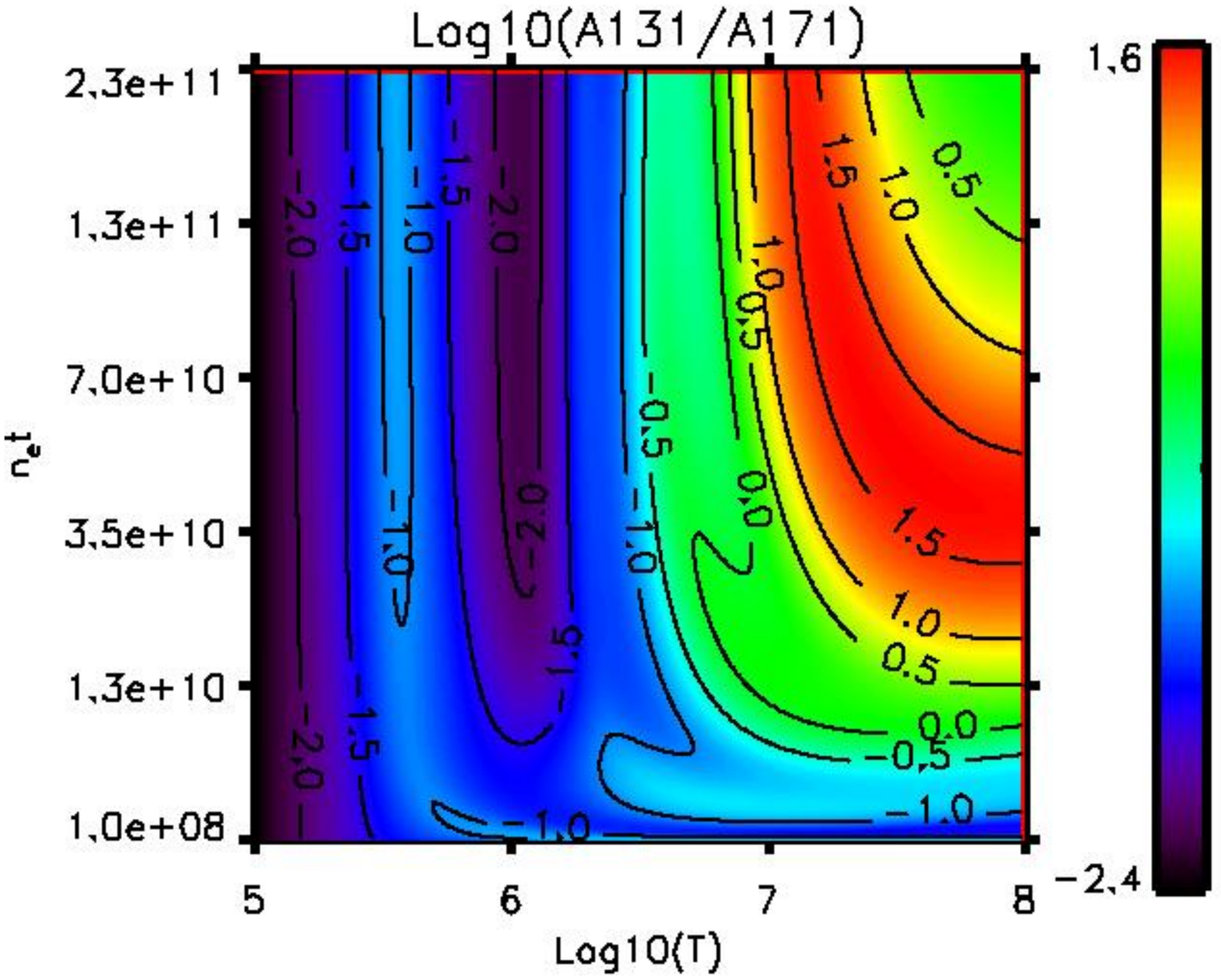}
\includegraphics[width=80mm]{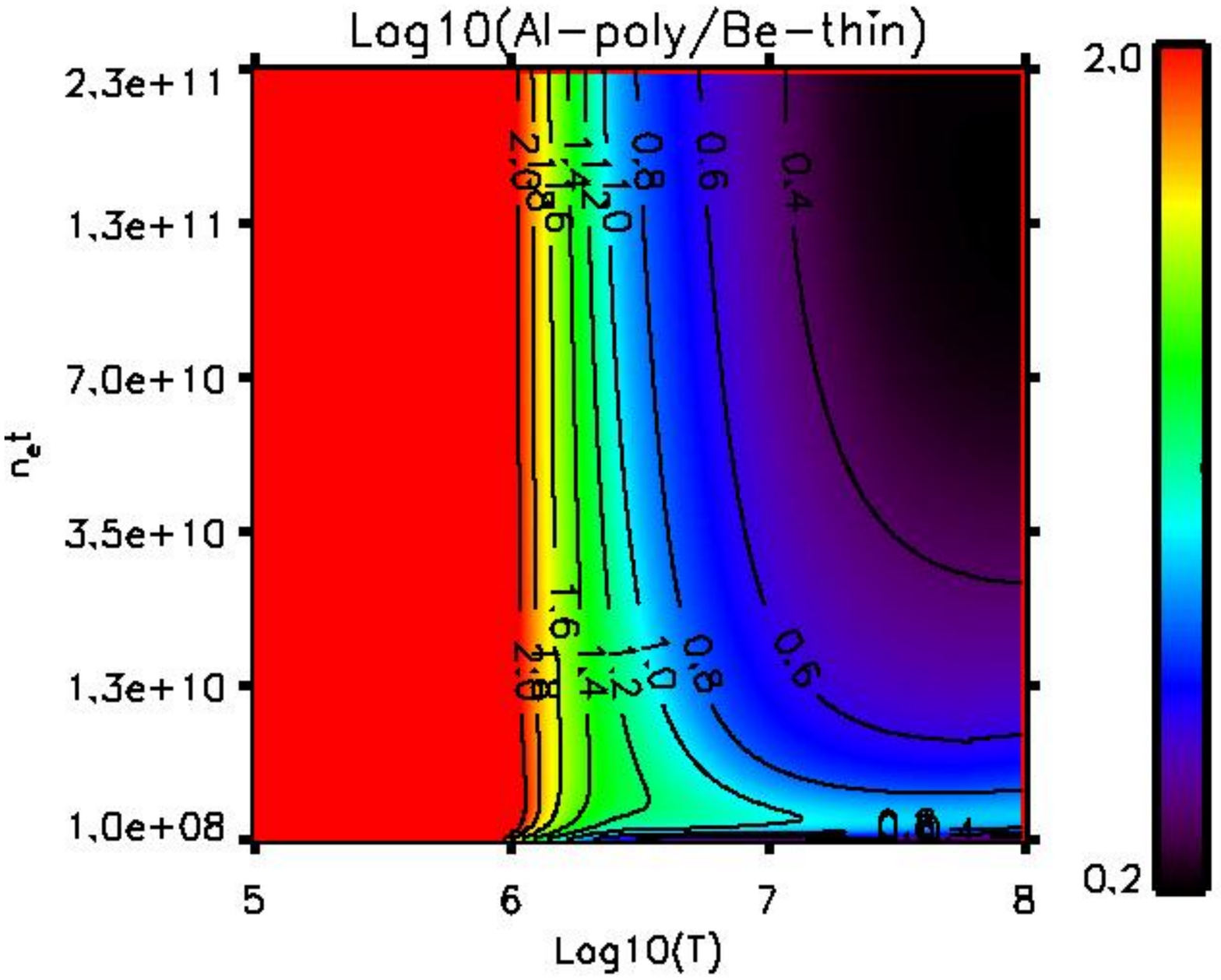}
\caption{131~\AA/171~\AA\ and Al\_poly/Be\_thin in temperatures and characteristic timescales 
in the left and right panels, respectively. 
In the right panel, the bigger ratios than 2.0 are represented by red colors below LogT=6.0. 
The ratios are shown with contours. }
\label{fig:2dr}
\end{figure}

We examine the two dimensional response ratios as a function of 
temperature and $n_{e}t$, which are calculated in Section 2.3, 
and examine whether they can be used to constrain temperature and density using the AIA and XRT observations. 
As an example, we show two passband ratios from AIA and XRT in Figure~\ref{fig:2dr}. 
We show a combination of 131~\AA\ and 171~\AA\ in the left panel of Figure~\ref{fig:2dr}. 
The ratio is a good indicator to determine 
the existence of hot plasma.  
The AIA ratio is almost independent of $n_{e}t$ at low temperatures 
because similar low ionization species, O~V, O~VI, Fe~VIII, dominate both bands. 
At temperatures above about 10$^7$~K, a strong dependence of the ratio on $n_{e}t$ is apparent 
due to the time needed to ionize into and out of the Fe~XX and Fe~XXI ions 
that are found near 10$^7$~K in equilibrium. 
A filter ratio between the XRT passbands has been also used to 
estimate the temperature of the observed plasma \citep[e.g.][]{lee2015}. 
We show a combination of Al\_poly and Be\_thin in the right panel of Figure~\ref{fig:2dr}. 
The XRT ratio is dominated by the Boltzmann factor, $e^{-hv/kT}$, 
so it is primarily dependent on T. 
Both plots of AIA and XRT show a banana shape with the ratios of at about -0.5$\sim$0.5 (red and green) 
in the left panel and $<$~0.8 (blue and purple) in the right panel 
at higher temperatures than $\sim$~LogT=6.5 in  the AIA and XRT, respectively. 
The ratios are not different with various $n_{e}t$ values for most temperatures. 
Therefore, it is hard to constrain the temperature and density with these plots. 

We try another method, a ratio-ratio plot. 
As examples of ratio-ratio plots, we show three ratio-ratio plots in Figure~\ref{fig:rr} for several temperatures and values of $n_et$. 
The different symbols and colors indicate the different temperatures and values of $n_{e}t$, respectively. 
The red solid curves are the ratio-ratio values corresponding to equilibrium. 
The red symbols indicate the corresponding equilibrium temperatures along the red curve. 
If any observed value lies on a straight line between points on the red curve, 
plasma with a combination of temperatures in equilibrium can match the observation. 
For example, the temperature at the location of `A' (black $\times$) in the black dashed line in the top panel of Figure 6 
can be interpreted as a combination of  3.2~MK (red star) and 13~MK (filled red diamond) plasmas in equilibrium. 
In that case the contributions of 3.2~MK and~13 MK can be found with a relation of $a:b=(13MK-3.2MK):(T_\times-3.2MK)$, 
where $a$ and $b$ are the distances along the black dashed line between two ratio-ratio points and T$_\times$ is the temperature at the point~$\times$. 
We find that the temperature is 6.2~MK with the relation. 
The temperature of `A' can also be estimated by the combination of temperatures of 4~MK (red triangle) and 20~MK (filled red square) on the black dotted line, 
In this case, the temperature is 18~MK. However, many other combinations of temperatures could produce the observed ratios.
As another example, the temperature at the location of `A$'$' (black +) on the black dash-dotted line 
could be an average temperature of 8.7~MK produced by a combination of 3.7~MK and 13~MK equilibrium plasmas. 
Otherwise, if the observed value does not lie on a line between points on the red curve 
(for example, the location of `B' in the top panel of Figure~\ref{fig:rr}), 
it cannot be produced by any combination of equilibrium plasmas, and 
we can get a rough idea of the density and temperature in NEI. 
Therefore, the ratio-ratio plot allows us to estimate the temperature using the combination of temperatures in equilibrium and 
give a rough idea how far the observed plasma is from EI in those cases. 
We show the region where the temperature can be estimated by some combination of temperatures in equilibrium in yellow in Figure~\ref{fig:rr}. 

\subsection{Application to the post-flare arcade in 2012~January~27}
\begin{figure}
\centering
\includegraphics[width=80mm]{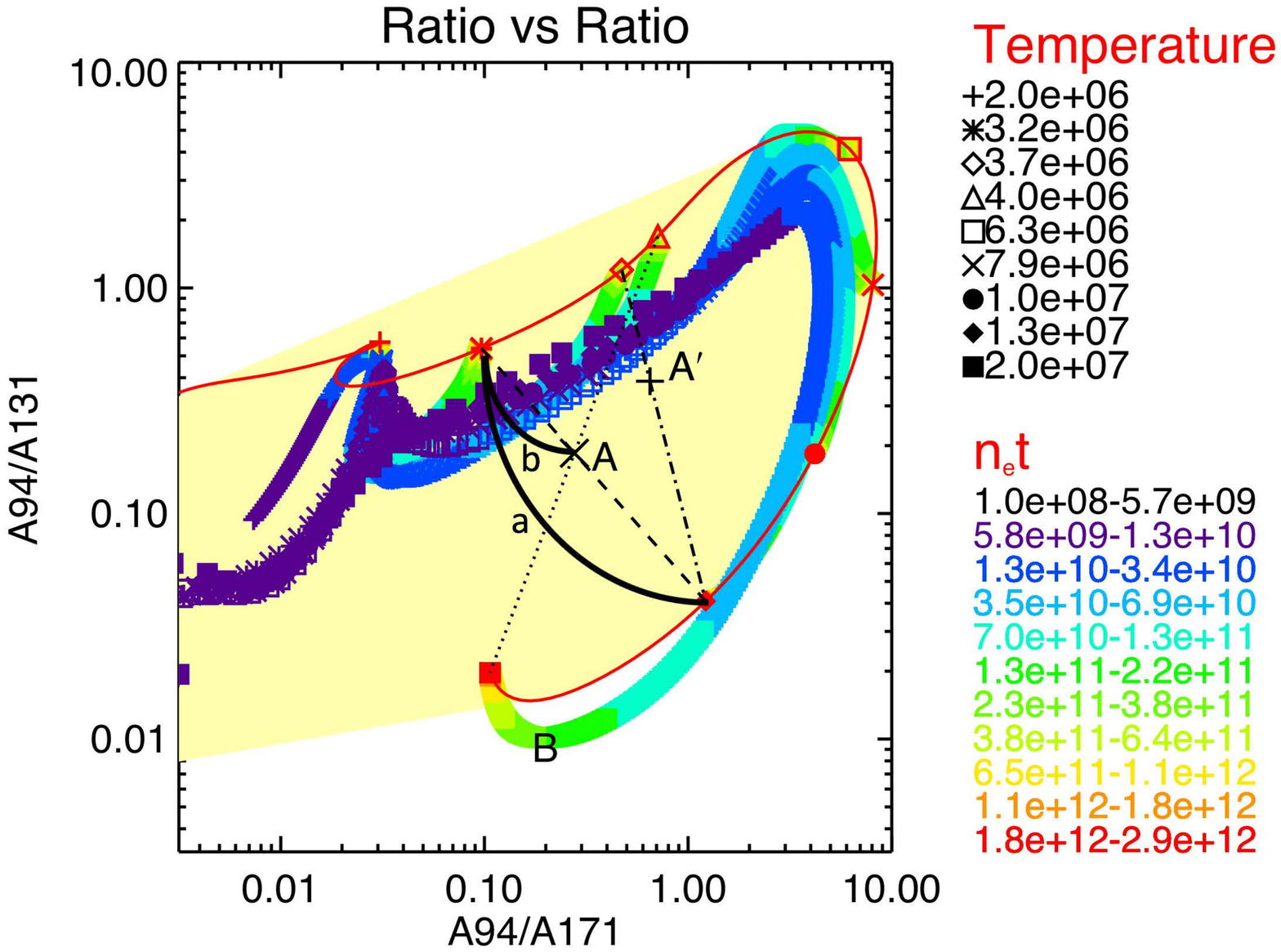} \\
\includegraphics[width=80mm]{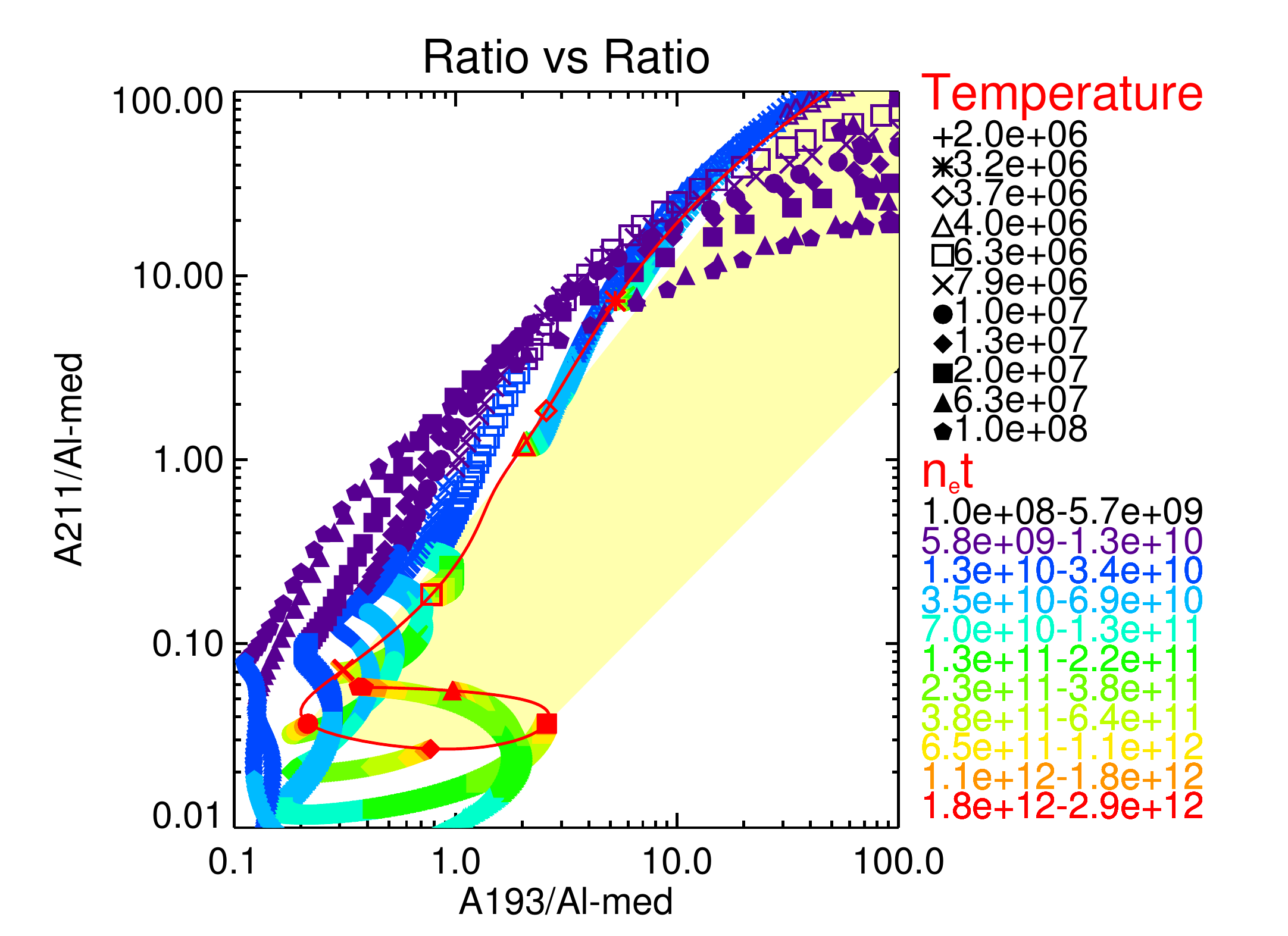} \\
\includegraphics[width=80mm]{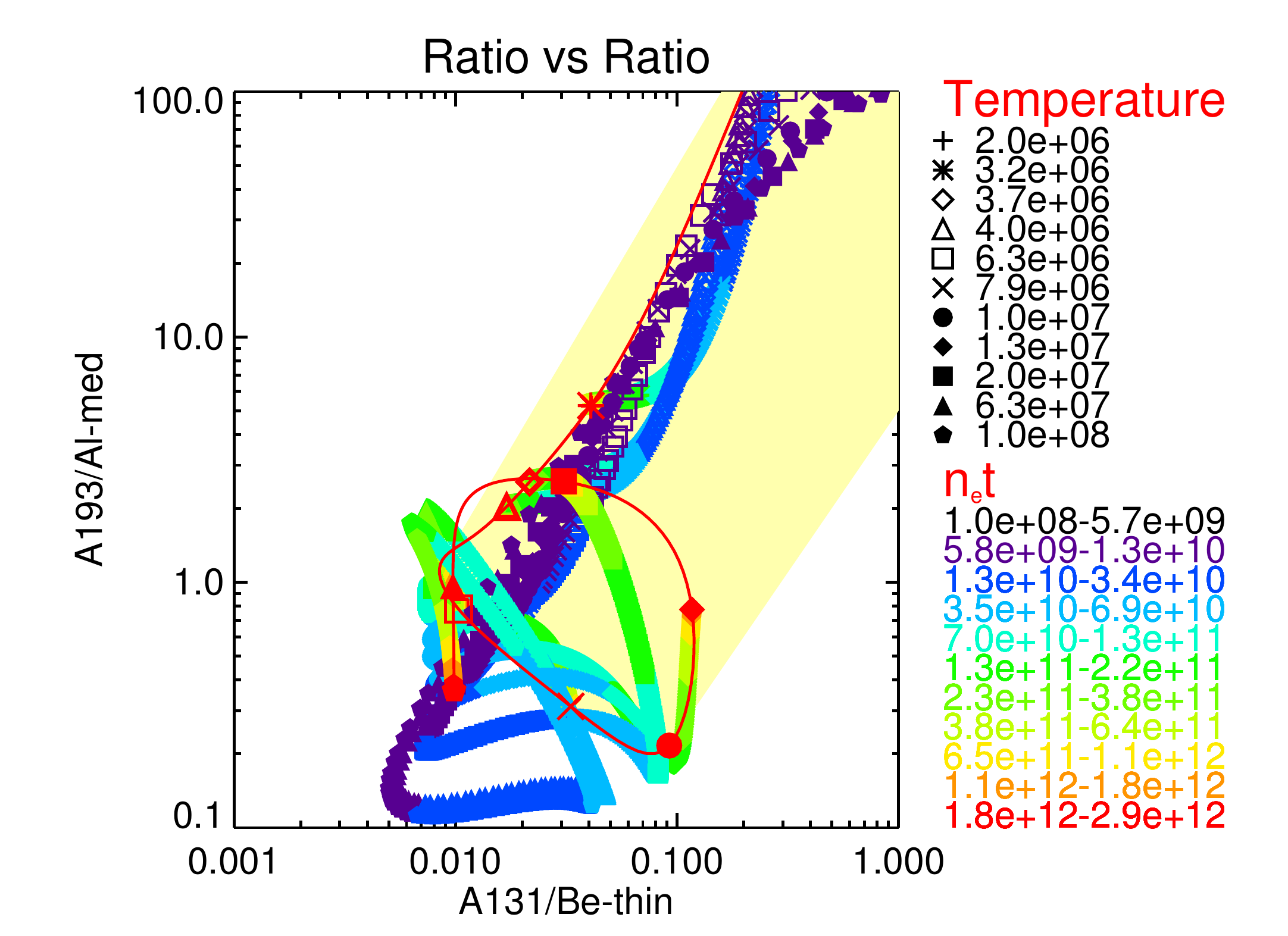} 
\caption{Ratio-Ratio with different temperatures (symbols) and characteristic timescale (colors) of 
94~\AA/171~\AA\ vs. 94~\AA/131~\AA\~(top), 193~\AA/Al\_med vs. 211~\AA/Al\_med~(middle), and 131~\AA/Be\_thin vs. 193~\AA/Al\_med~(bottom). 
We show the temperature range from 1~MK to 20~MK in the top panel. 
Red solid curves with symbols represent the ratio-ratio in equilibrium. 
Please see text for `A', `A$'$', `B', a, and b in the top panel. 
Yellow represents the region where the temperature can be estimated by some combination of temperatures in equilibrium.}
\label{fig:rr}
\end{figure}

\begin{figure}
\centering
\includegraphics[width=150mm]{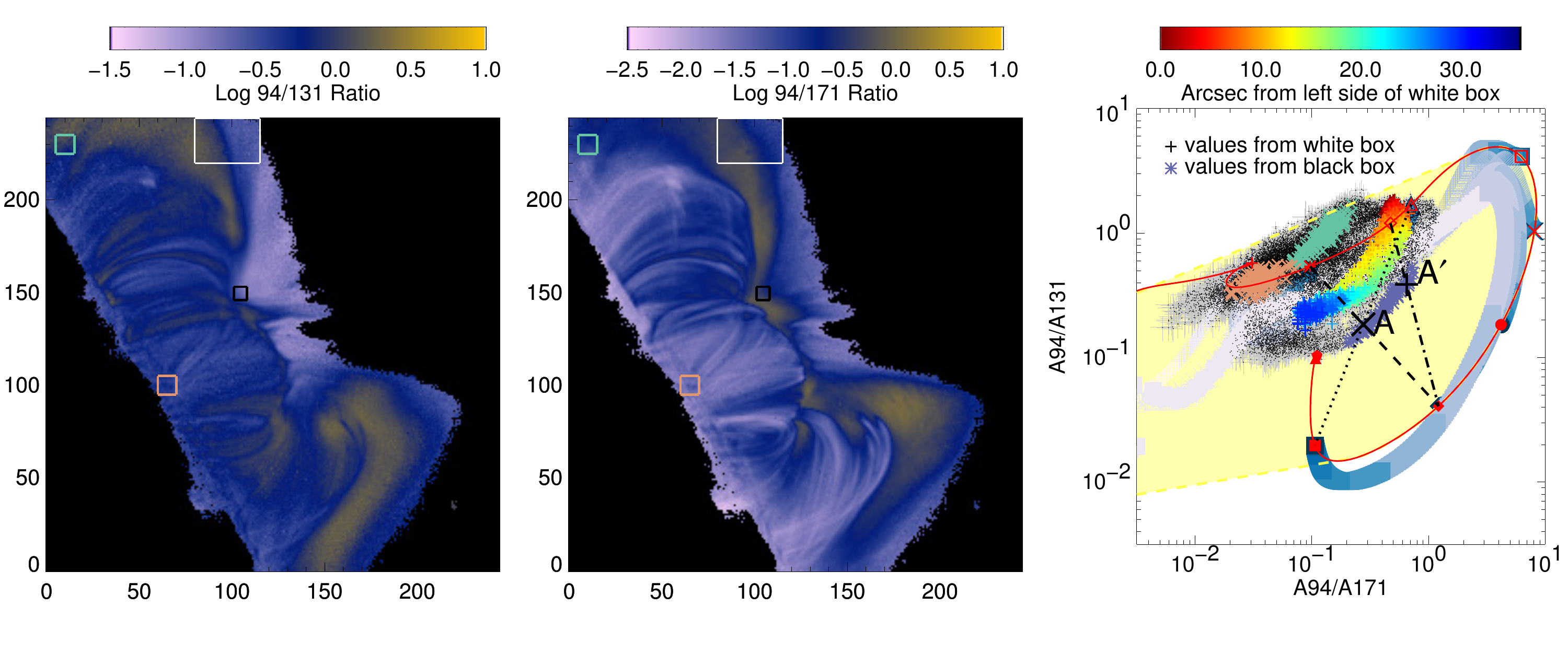} \\
\includegraphics[width=150mm]{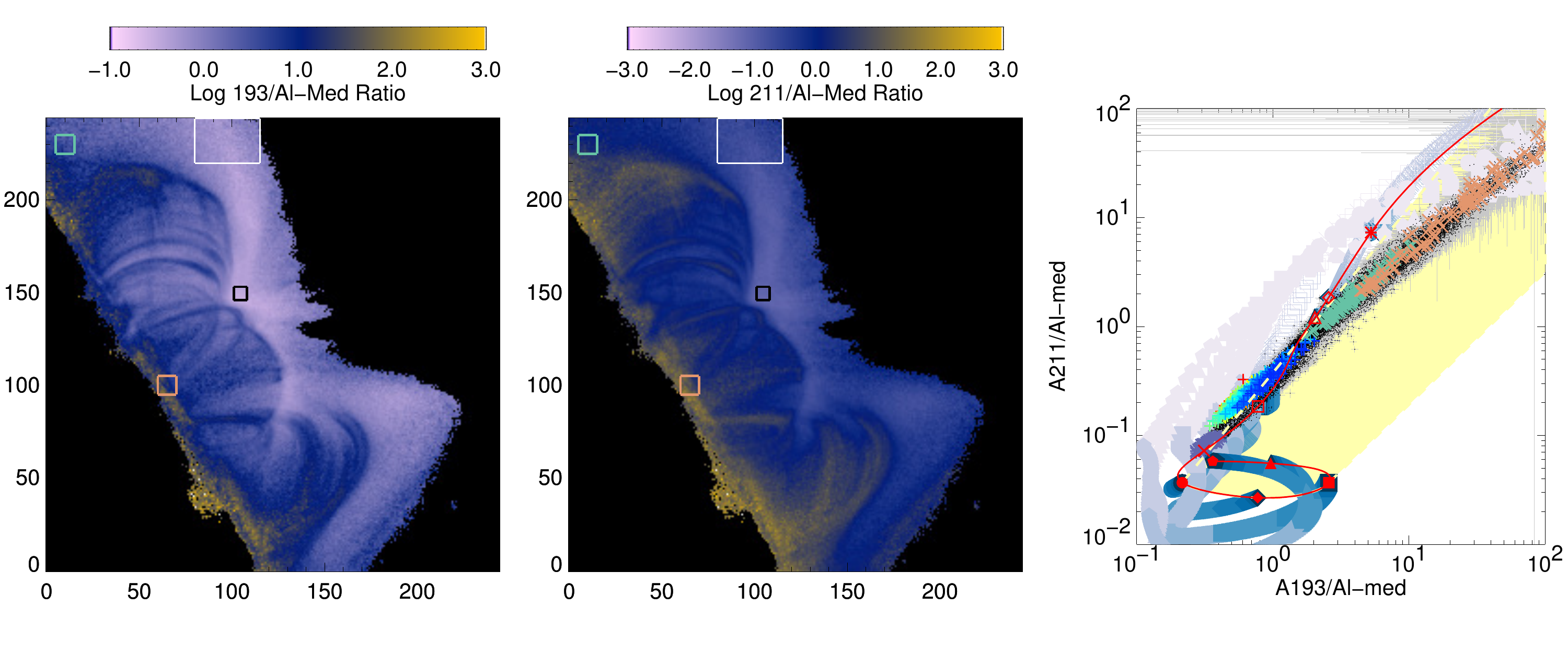} \\
\includegraphics[width=150mm]{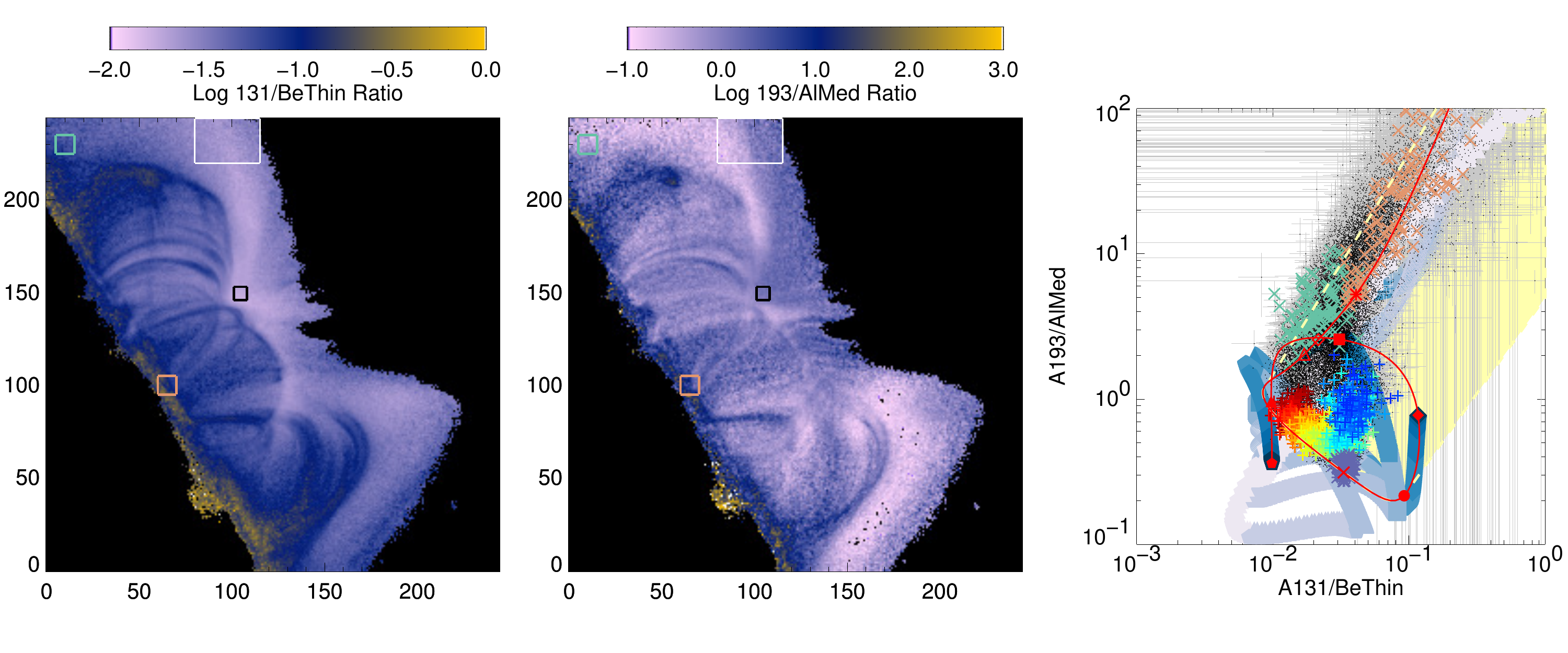}
\caption{ Post-flare loop arcade at 20:45 UT on 2012 January 27. Ratio images of 94~\AA /131~\AA\ and 94~\AA /171~\AA\ (top), 
93~\AA /Al\_med and 211~\AA /Al\_med (middle), and 131~\AA/Be\_thin and 193~\AA/Al\_med (bottom) and their ratio-ratio plots. 
Black dots on the ratio-ratio plots represent the ratios for pixels on the ratio images. 
The ratios of the pixels within the white box on the left panels are represented by rainbow colored crosses in the right panels. 
The color indicates the distance from the left side of the white box. 
The ratios of the pixels within black boxes on the left panels are represented by purple stars in the right panels. 
Pastel blue and orange $\times$ represent the ratios within the pixels within the same colored boxes. 
On the ratio-ratio plots, the red curves and the red symbols are the same as in Figure~\ref{fig:rr}. 
Grey bars represent uncertainties in both ratios for each point on the plot. 
Ratios with a value of 100 or greater involving the Al-med filter should probably not be believed. 
The lower left of the swarm of points in the 94~\AA/171~\AA\ vs. 94~\AA /131~\AA\ plot is more uncertain, because there are more low value 94~\AA\ points there. 
Yellow regions are the same as in Figure~\ref{fig:rr}.} 
\label{fig:rimg}
\end{figure}

Figure~\ref{fig:rimg} shows the ratio images of a post-flare loop arcade associated with an X1.7 flare 
observed by AIA and XRT on 2012 January 27. 
We discuss 
whether the observed ratios are consistent with ionization equilibrium by comparing them with the model ratios in Figure~\ref{fig:rr}. 
Black dots in the right panels are the ratios of all pixels in the left ratio images. 
Grey bars are uncertainties in both ratios for each point calculated by using a formula in \citet{lee2017}, 
which tends to Gaussian for high count regimes and Poissonian for low count regimes \citep{gehrels1986}. 
We show the model ratios in Figure~\ref{fig:rr} with pastel blue colors to help compare the observations with the model. 
Rainbow colored crosses and purple stars in the right panels are the ratios of the pixels within the white box and the black box, respectively, in the left panels. 
Pastel orange and green colors are the ratios of the pixels within the pastel orange box near foot points of loops and 
the pastel green box on outer larger loops, respectively. 
The observed ratios near the foot points (pastel orange color) are close to the equilibrium temperatures between 2~MK (red cross) and 3.2~MK (red star). 
In the top right panel, the cloud of black dots above the red curve near [0.1, 0.5] requires a combination of equilibrium components 
with temperatures between 2~MK and 4~MK,  
and these ratio values are not able to be explained by NEI because the NEI solutions all lie below the red curve (see Figure 6). 
The observed points from the pastel green box are located in this cloud of black dots, 
indicating that these loops are mute-thermal and have temperatures between 2 MK and 4 MK.  
Yellow regions are the same as in Figure~\ref{fig:rr}. 

In this ratio-ratio plot with AIA only pairs, the swath of rainbow colored crosses agrees very well with the model ratios 
for 3.7~$-$~4~MK (diamond and triangle symbols) as the $n_et$ values are varied. 
The rainbow colored crosses move from red to blue as their associated location moves further from the left hand side of the white box. 
Thus the location of these values in the ratio-ratio plot is consistent with plasma at 3.7$-$4~MK that is closer to equilibrium the closer it is to the arcade. 
However, we note that we cannot rule out that the plasma in the white box is a result of a combination of plasma in equilibrium at different temperatures, 
since the rainbow colored crosses are within the red equilibrium curve. 

In the middle right panel, the black dots with 193~\AA/Al\_med between 1 and 3 require one component
in a narrow temperature range between 3.7~MK and 6.3~MK  in equilibrium. 
The black dots with 193 A/Al med $>$ 3 require a combination of temperatures between 2 and 6 MK.  
These points are all located at the footpoints of the flare loops (pastel orange color), 
so they probably correspond to pixels with more than one temperature along the line of sight.
The black dots with 193~\AA/Al\_med $<$ 1 require a combination of two components of 6.3~MK and 10~MK in equilibrium, but probably an isothermal 6 MK plasma is 
within the uncertainties, or else a hotter plasma (e.g., filled circle, diamond, and square) with moderate $n_et$ (blue color in the middle panel of Figure~\ref{fig:rr}). 
The rainbow colored crosses can be explained with a combination of 4~MK and 7~MK in equilibrium. 
The red and yellow crosses are hard to see
here because these are underneath the green and blue crosses.
This result indicates higher temperatures than those seen in the top panel. 
We note that combinations between 4 MK and 7~MK would not work to explain the rainbow colored crosses in the top right panel. 
The pastel orange ($<$3.7~MK) and green points (between 3.2~MK and 4~MK) also show higher temperatures than the temperatures (2~MK and 3.2~MK) in the top panel.  
One possibility for the discrepancy could be that the 131 \AA\ and 171 \AA\ filters are both sensitive to plasma at around 1 MK, 
which may be contaminating the points in the top ratio-ratio plot. 

In the bottom right panel, the black dots with 193~\AA/Al\_med $>$ 3 can be explained with a single low temperature (e.g., 3.2~MK) in equilibrium. 
The black dots with 193~\AA/Al\_med $<$ 3 require a combination of temperatures in equilibrium, which are mostly inside the red curve. 
The red and yellow crosses are close to equilibrium, but the temperatures are between 6.3~MK and 8~MK. 
The ratios of green and blue crosses might come from the plasmas between 4~MK and 8~MK. 
These temperatures are higher than the ones in the top panel but similar to the temperatures in the middle panel. 
The pastel orange and green points are also similar to the temperatures in the middle panel. 

\citet{hanneman2014} have calculated a differential emission measure (DEM) for the supra-arcade region observed by AIA and XRT at the same time as in Figure~\ref{fig:rimg} assuming EI. 
From the DEMs, they find that there are three temperature components of about 1~MK, 6$-$8~MK and 30 MK for the flare arcade (`ARC' in Figure~20 in their paper). 
The low temperature plasma of 1~MK is possibly from background or foreground but not likely from the arcade. 
It is possible that this low temperature plasma affects the AIA only pair ratios, but has less of an effect the other AIA and XRT pair ratios. 
We select a similar location with the arcade near [100, 150] (black box). 
The ratios are seen as purple stars in the ratio-ratio plots. 
The locations of `A' and `A$'$' in Figure~\ref{fig:rr} correspond to the ratios of the purple stars for the observed points within the black box. 
We show the `A' and `A$'$' also in Figure~\ref{fig:rimg}.  
The average temperature of these points could be 6.2~MK or 18~MK on the location `A' and 8.7~MK on the location `A$'$' as estimated in Section 3.1. 
In the middle panel, the purple stars are in the head of a narrow long black cloud.  
This location in the ratio ratio plot is consistent with hot plasma about 8~MK (red $\times$) in EI or hotter plasma close to 20~MK (blue filled square 
in the bottom panel of Figure~\ref{fig:rr}) with the moderate $n_et$. 
In the bottom panel, the locations of the purple stars are consistent with about 8~MK (red $\times$) in EI or 20~MK (blue filled square) with moderate $n_et$, 
although these are overlapped with much higher temperatures (blue filled up and down triangles) which can be rejected by comparing with the plot in the middle panel. 
The two hot temperature components in the middle and bottom panels are similar to each other. 
The lower temperature component in EI is similar to the results in \citet{hanneman2014}. 
However, the temperature of 20~MK for the hotter plasma is lower than the temperature of 30~MK calculated by assuming equilibrium in \citet{hanneman2014}. 
In this case, the hotter temperature of 18~MK estimated from the AIA only pair is similar to the temperature of hotter plasma from the AIA and XRT pairs. 
The method of the ratio-ratio plot can give several temperatures from a ratio-ratio pair. 
Thus, we should consider several combinations of the ratio-ratio pairs, and find which solution might be reliable. 
It would be good to compare this method to many other events. 

Using the example of an application to the post-flare arcade, we find indications in the data that the plasma closer to the arcade is closer to EI. 
However,  we also find that most of the observed points may be described by using a combination of temperatures in equilibrium, 
so the presence of plasma out of equilibrium is difficult to establish definitively. 
In this example, the ratio-ratio plot with AIA only pairs gives lower temperatures than the temperatures in AIA and XRT pairs. 
It is possible that this discrepancy is because AIA is less sensitive to the hotter plasma while XRT is more sensitive to it. 
The ratio-ratio plot gives several different temperatures in EI or/and NEI. 
It is possible since the observed coronal plasma is multi-thermal rather than isothermal, and also there is an effect of the background emission along the line of sight. 
Foreground or background contributions will tend to pull the ratios inside the regions 
where combinations of equilibrium plasmas can account for the ratios. 
It would be best to subtract the background emission for comparison with the ratio-ratio plots, but that can be difficult if the background varies. 
Therefore, we should examine carefully several ratio-ratio pairs together. 
We show the first application of our NEI models to the observations. 
In the future, more detailed analysis is required with other observations. 

In this analysis, we find that the temperatures estimated from ratio-ratio plots 
using a combination of the AIA and XRT passbands are higher than 
the temperatures estimated by using ratio-ratio plots that only use the AIA only passbands. 
One possibility is that the 131~\AA\ and 171~\AA\ are less sensitive to the higher temperatures, 
so the passband pairs tend to the lower temperatures. 
However, we can not rule out a possibility of the calibration issue between the AIA and XRT instruments. 
If we adopt the factor of two multiplied to the calculated XRT responses for NEI \citep{wright2017, testa2011, cheung2015, schmelz2015}  
then the model ratios tend towards the lower left in the ratio-ratio plots. 
In this case, the temperatures obtained with ratio-ratio plots using both the AIA and the XRT passbands 
are lower and therefore more similar to the temperatures obtained with ratio-ratio plots using AIA only.  
The effect of the cross-calibration between the two instruments will need further investigation. 

It is hard to say exactly whether the observed ratios represent that the plasma is in EI or NEI. 
One possibility is that there are no observations that are certain to be in equilibrium. 
\citet{bradshaw2011} show that small-scale, impulsive heating including a nonequilibrium ionization state predicts the observable quantities 
that are entirely consistent with what is actually observed. 

\subsection{Application to the shocked plasma in 2010~June~13}

\begin{figure}
\centering
\includegraphics[width=70mm]{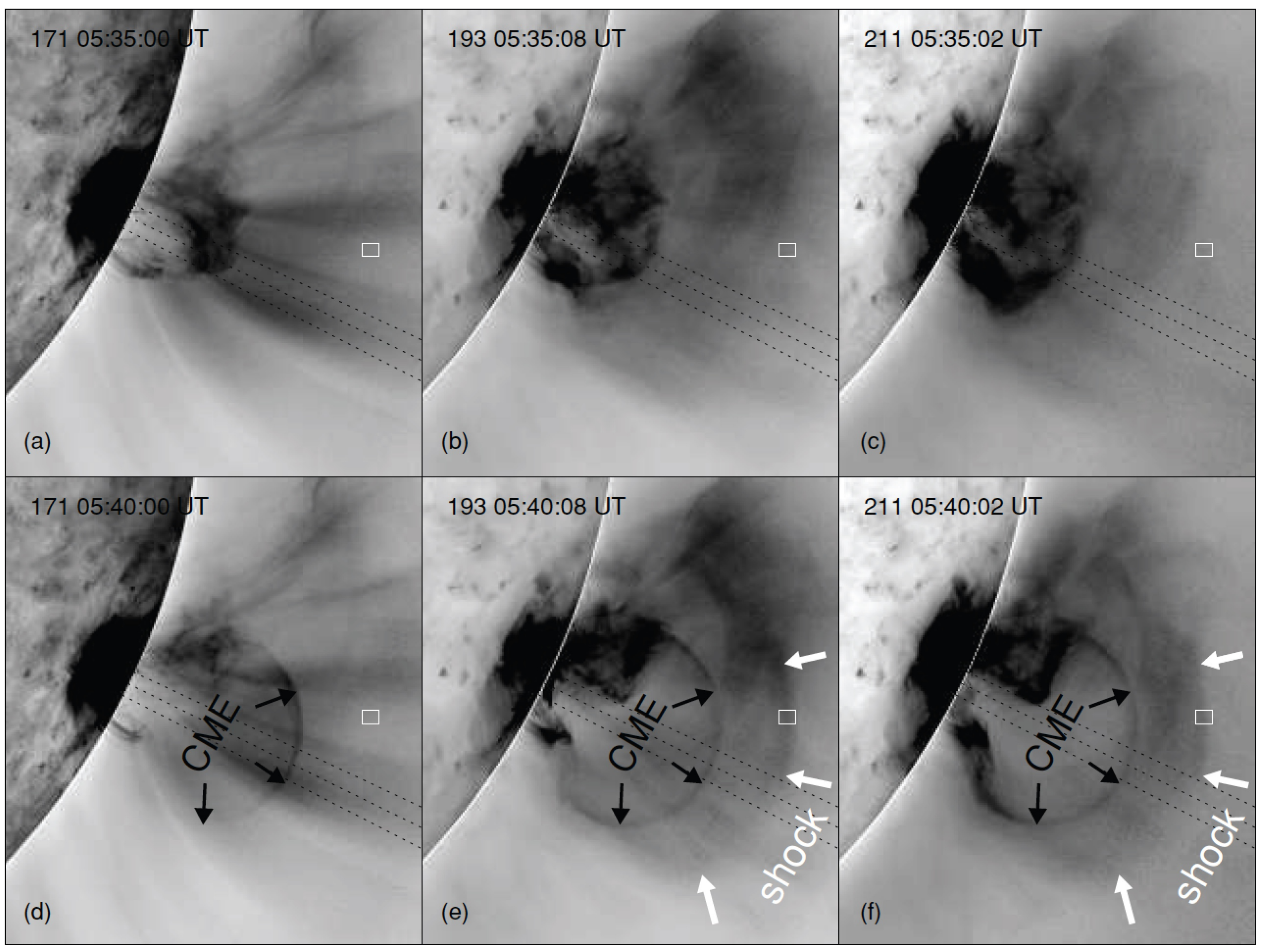}
\includegraphics[width=80mm]{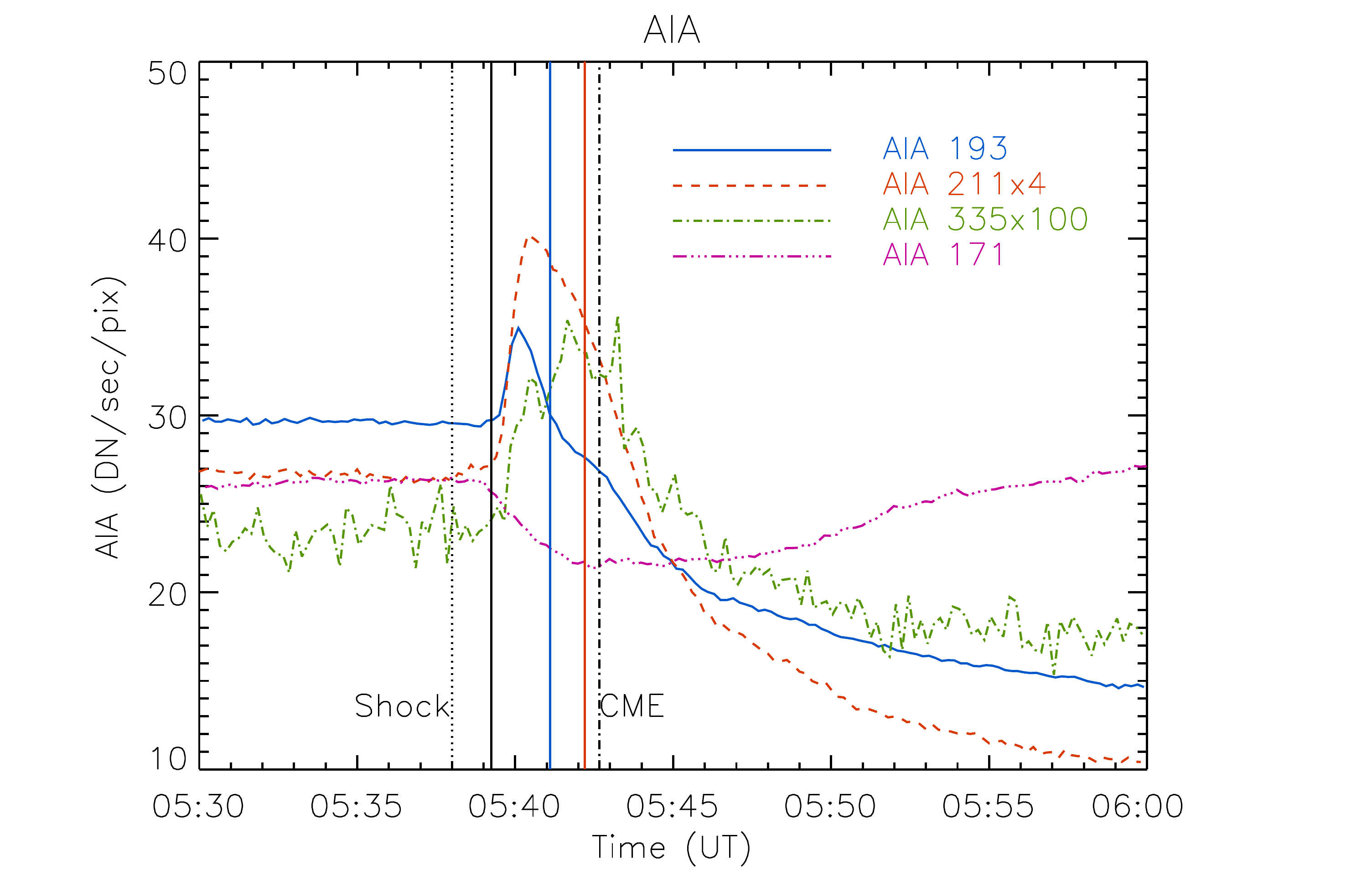}
\caption{Left: AIA observations showing the morphology of the shock wave (reversed color table). 
Figure is taken from \citet{ma2011}. 
Right: Intensity flux tracking in the white box in the left. 
The dotted and dash-dotted vertical lines represent the time when the shock and CME bubble arrived in the white box, respectively. 
See details \citet{ma2011}. 
Solid vertical lines are the start and end times for the comparison with the characteristic responses. 
Black is the start time ($\sim$05:39:15 UT) for 193\AA, 211\AA, and 335\AA, 
blue is the end time (05:41:06 UT) for 193 \AA, and red is the end time ($\sim$05:42:15 UT) for 211\AA\ and 335\AA. }
\label{fig:shock}
\end{figure}

We apply our results to the CME-driven shock of 2010 June 13 studied by Kozarev et al. (2011) and Ma et al. (2011).
We use the count rates in the AIA 193~\AA\, 211~\AA\ and 335~\AA\ bands measured in the white box 
in Figure \ref{fig:shock} by Ma et al. (2011). 
In this event, the ratio plots (Figure~\ref{fig:2dr}) and ratio-ratio plots (Figure~\ref{fig:rr}) with the three passbands are hard to 
apply directly to distinguish the temperature and density, because each ratio corresponds to a banana-shaped region
in the T-$n_{e}t$ plane. For this reason, we match the observed intensity histories to the
characteristic timescale responses and find the temperature, density, and line of sight depth ($\it{dl}$) ranges that satisfy the observations. 
The advantage of this method is that it incorporates the information in the time dependence of the shocked plasma.

The method assumes that the plasma is in EI before the shock and, then the plasma within the
white box was shocked at earlier times as the shock passes beyond the box.  We use the averaged intensities 
between $\sim$05:30:00~UT and $\sim$05:39:00~UT for each passband as the pre-shock background. 
As the shock bubble expands past the box, more and more of the background plasma is pushed out in front of the shock wave.  
The amount of gas in the white box is constant or slightly increased
because of the shock compression.  If the Fe IX ionization fraction
stayed constant, the 171 \AA\ band would brighten.  Therefore, we model
the light curve under the assumption that the Fe IX is ionized away in
the pre-shock gas, so that the fading in the 171~\AA\ band tracks the reduction of the background. 
We then multiply the pre-shock 
backgrounds in the other bands by the ratio of the 171 \AA\ band count rate to the pre-shock 171 \AA\ band value. 
We indicate the start and end time of the observations used for the comparison with the characteristic responses in the right panel of Figure~\ref{fig:shock}. 
We use the 193~\AA\ observations from 05:39:18~UT to 05:41:06~UT. 
The 211~\AA\ and 335~\AA\ observations are used from 05:39:12~UT and 05:39:15~UT to 05:42:12~UT and 05:42:15~UT, respectively. 
This time range avoids the arrival (dash-dotted line in Figure~\ref{fig:shock}) of CME material within the box 
and only part of the box contains shocked plasma for a few exposure times of the observations.  
By dividing $n_{e}t$ by density, we have the characteristic responses as a function of time. 
Thus, we can compare the observations with the characteristic response in times once we give a specific density and a line of sight depth. 
We apply the grid of T from $2 \times 10^6 \rm~K$ to $10^8 \rm~K, n_e$ from $10^7 \rm~cm^{-3}$ to $10^9 \rm~cm^{-3}$, 
and $dl$ from $10^8 \rm~cm$ to $10^{11} \rm~cm$. 
We then search for combinations of T, $n_e$, and $\it{dl}$ that match the observed count rate curves. 

\begin{figure}
\centering
\includegraphics[width=80mm]{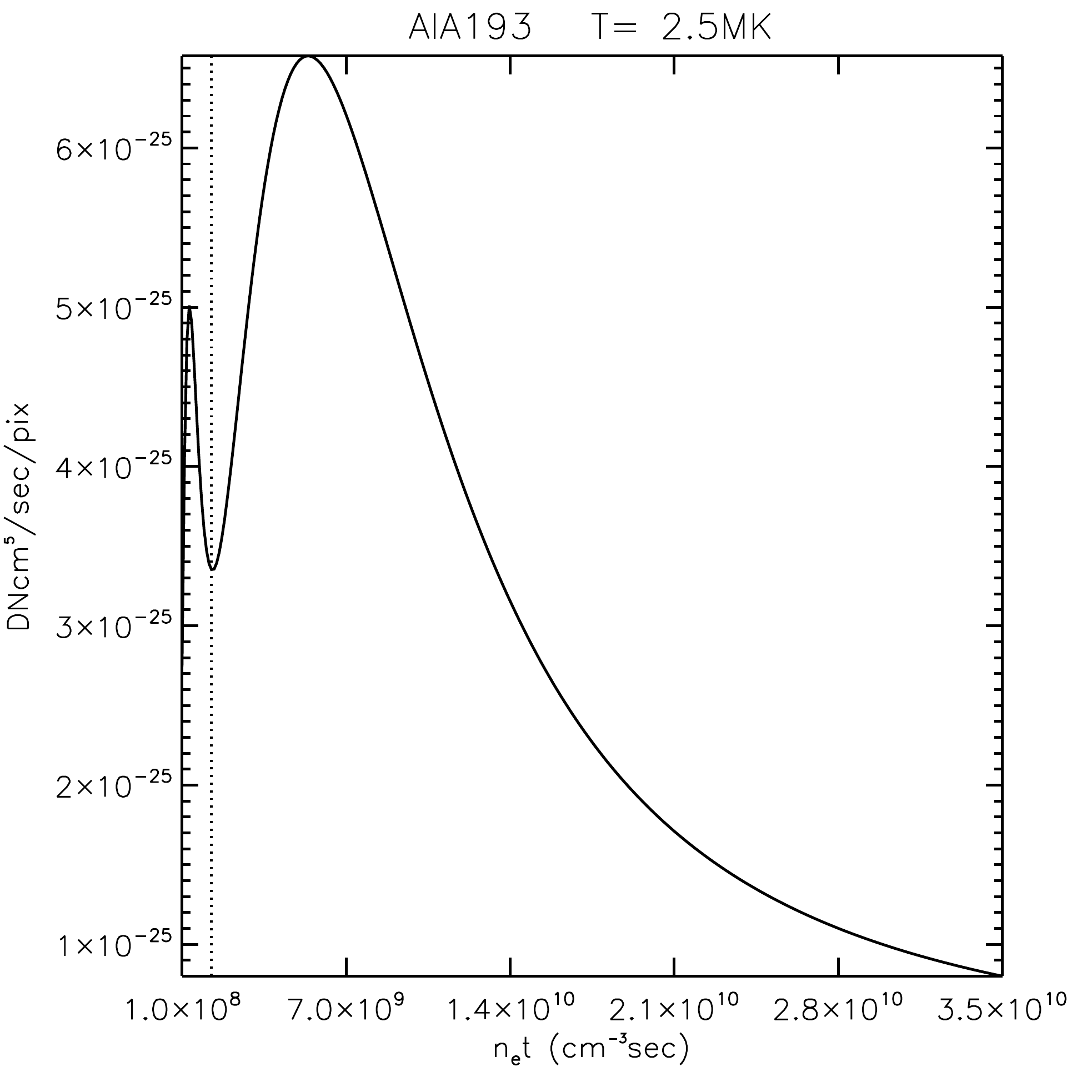}
\includegraphics[width=80mm]{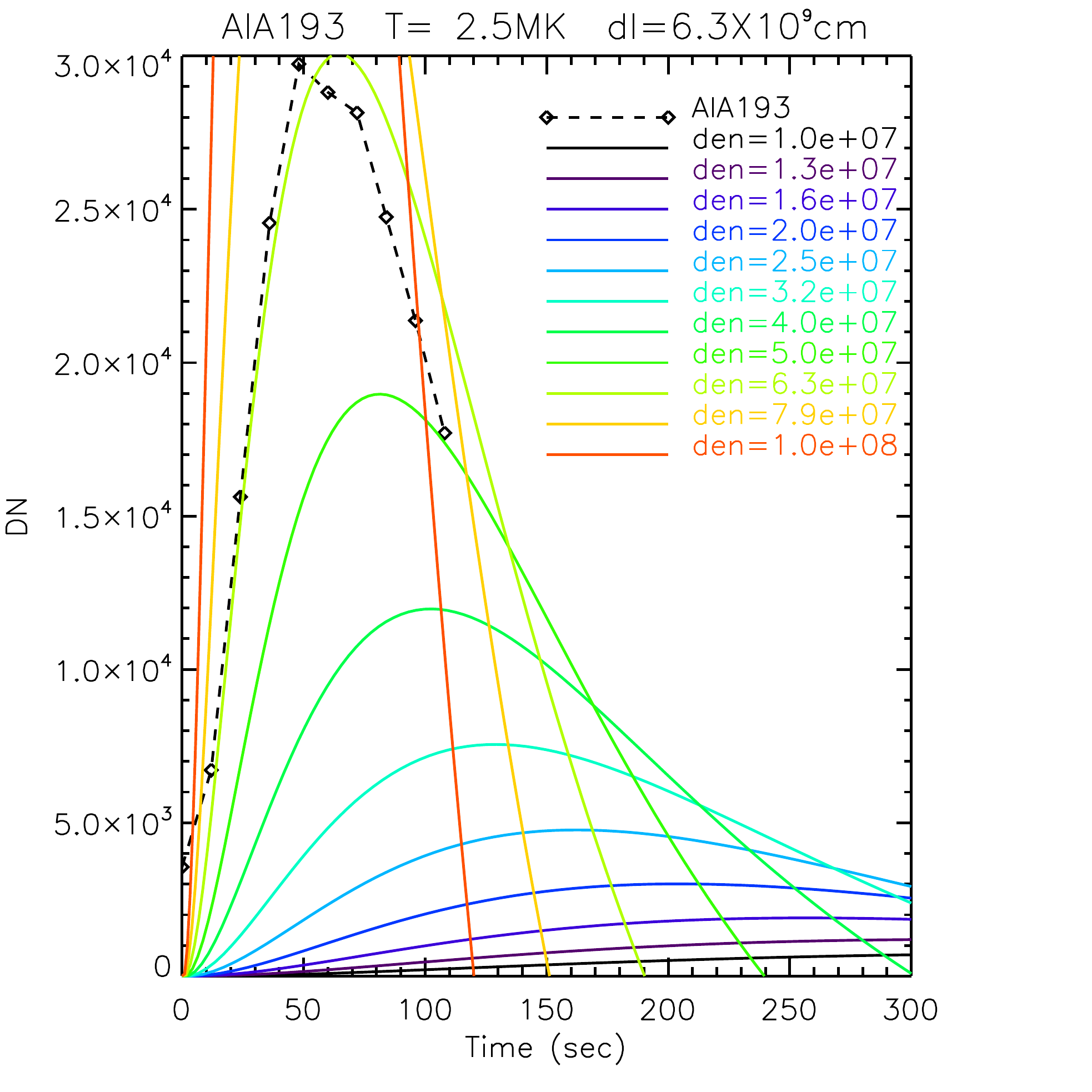}
\caption{Left: Characteristic time scale response of 193~\AA\ at 2.5~MK. 
Dotted line represents the starting $n_et$ of the response for the comparison with the observation. 
Right: Comparisons the observed intensity of 193~\AA\ (dashed black line with a diamond symbol) 
with the model responses with different densities by different colors.}
\label{fig:shock_resp}
\end{figure}

The left panel of Figure~\ref{fig:shock_resp} shows the characteristic response of the 193 \AA\ band at 2.5 MK as an
example. The second peak of the response near $n_et \sim 5 \times 10^9 \rm ~cm^{-3} ~sec$ is the major 
contribution of Fe XII in the 193 \AA\ passband as seen in Figure 2. At the very low $n_et$, the peak may be from Fe IX, 
and it is not the major contribution to 193~\AA\ band.  We compare the responses starting at the minimum shown by the 
dotted vertical line in the left panel in Figure~\ref{fig:shock_resp}. 
The emission at the minimum is subtracted from the response, and the shock emission is set to zero at the initial time. 
That is reasonable because the emission by shock at 193 A is
mostly due to the Fe XII and the emission at the minimum is due mostly to low temperature lines.
For the comparison with the observations in DNs, 
we multiply $n_e^2, dl$, exposure time, and the number of pixels to the characteristic response. 
In the right panel of Figure~\ref{fig:shock_resp}, we show the observations in DNs  
(dashed black line with a diamond symbol) and the responses for different densities, 
indicated by different colors, at T=2.5~MK and ${\it{dl}}=6.3 \times 10^9 \rm~cm$ 
as an example.  

We compare the observed profile of each passband with the responses for all T, $n_et$, and $\it{dl}$. 
Then, we find the allowed ranges of temperature, density, and line of sight depth where Chi-squared is less 
than it's minimum value + 1.6.  We use 1.6 rather than 1.0, because the RMS deviations in the pre-CME 
background DN levels were about 1.6 times larger than expected from the count rate alone.
The reduced $\chi^2$ is given by, 

\begin{equation}
\chi^2 = \frac {\sum_{t=1}^{n} {\sqrt{( I_{band} (t) - f~ n_e^2~ dl~ R(T, n_et, band) )^2}}/ {I_{obs, band}(t) } } {(n-1)}, 
\end{equation}

\noindent where f is a constant, exposure time $\times$ the number of pixels within the box (40$\times$32~pixels). 
The exposure times during the observations are 2.9 sec for all four passbands. 
I$_{band}(t)$ is the observed DNs after the background subtraction. 
I$_{obs, band}(t)$ is the observed DNs. The number of observations, n, is 10, 16, and 16 for 193 \AA\, 211\AA\, and 335\AA, respectively.  

\begin{figure}
\centering
\includegraphics[width=100mm]{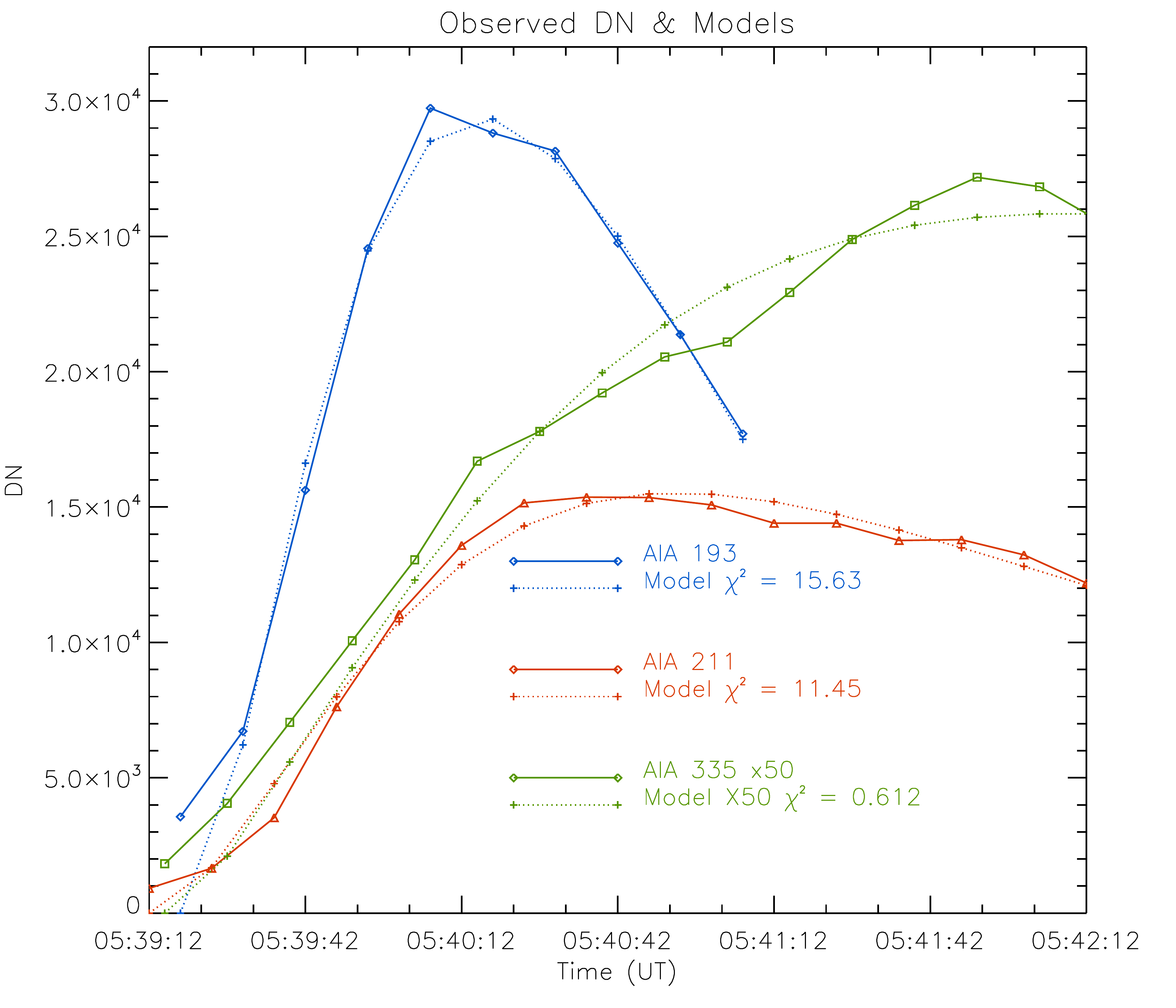}
\caption{Blue, red, and green solid lines represent the observed DN 
for 193 \AA, 211 \AA, and 335 \AA, respectively. 
Dotted lines are the model responses 
with the best fitting in the constraints 
in Figure~\ref{fig:shock_cons}.}
\label{fig:shock_profile}
\end{figure}

We show the profiles of observations and the predicted models in Figure~\ref{fig:shock_profile} with the reduced $\chi^2$ values. 
The dotted lines with a cross symbol are the best models. 
The reduced $\chi^2$ value for 335~\AA\ is smaller than others 
since the uncertainty of 335~\AA\ is relatively much larger than the observations in the 193 \AA\ and 211 \AA\ bands. 

\begin{figure}
\centering
\includegraphics[width=80mm]{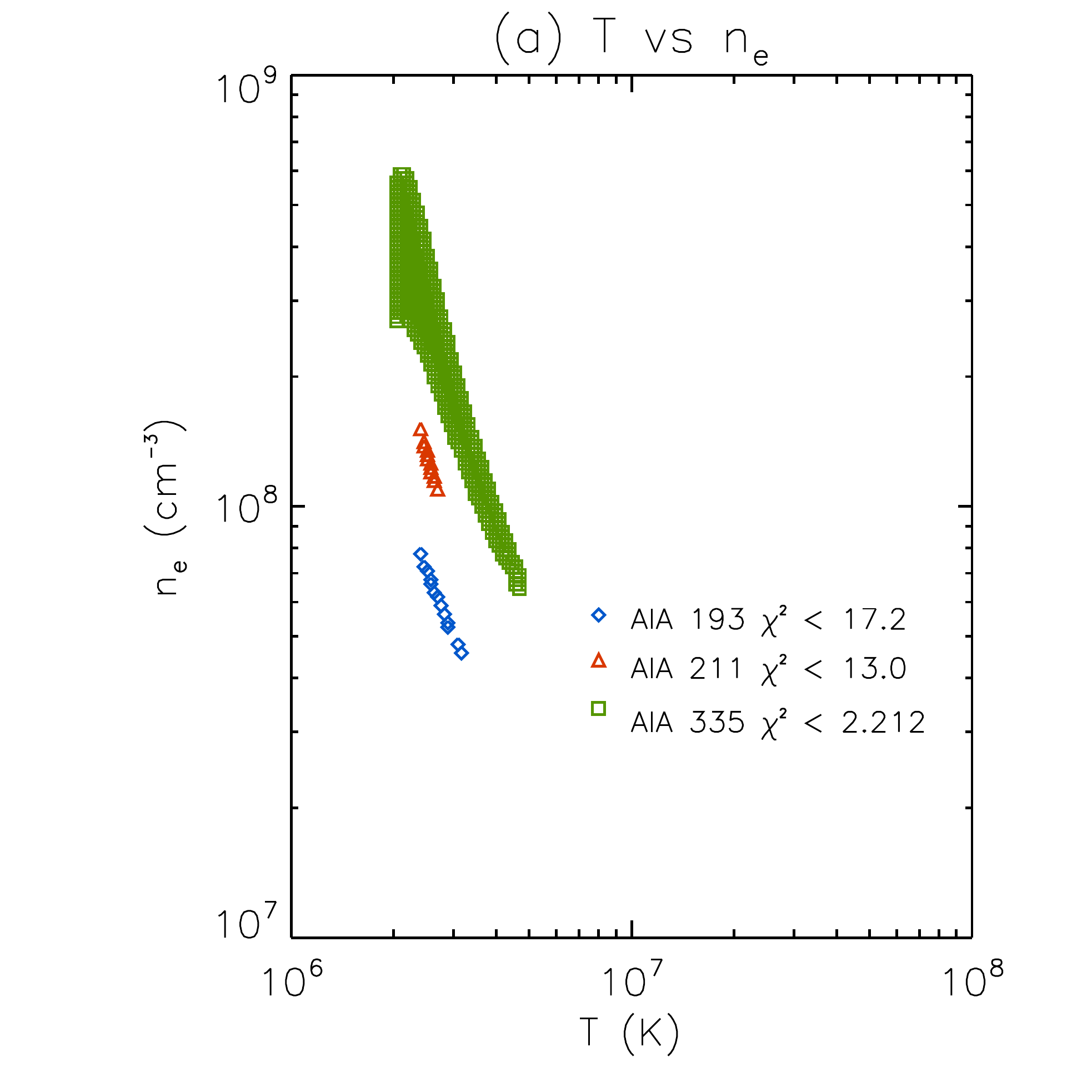}\\
\includegraphics[width=80mm]{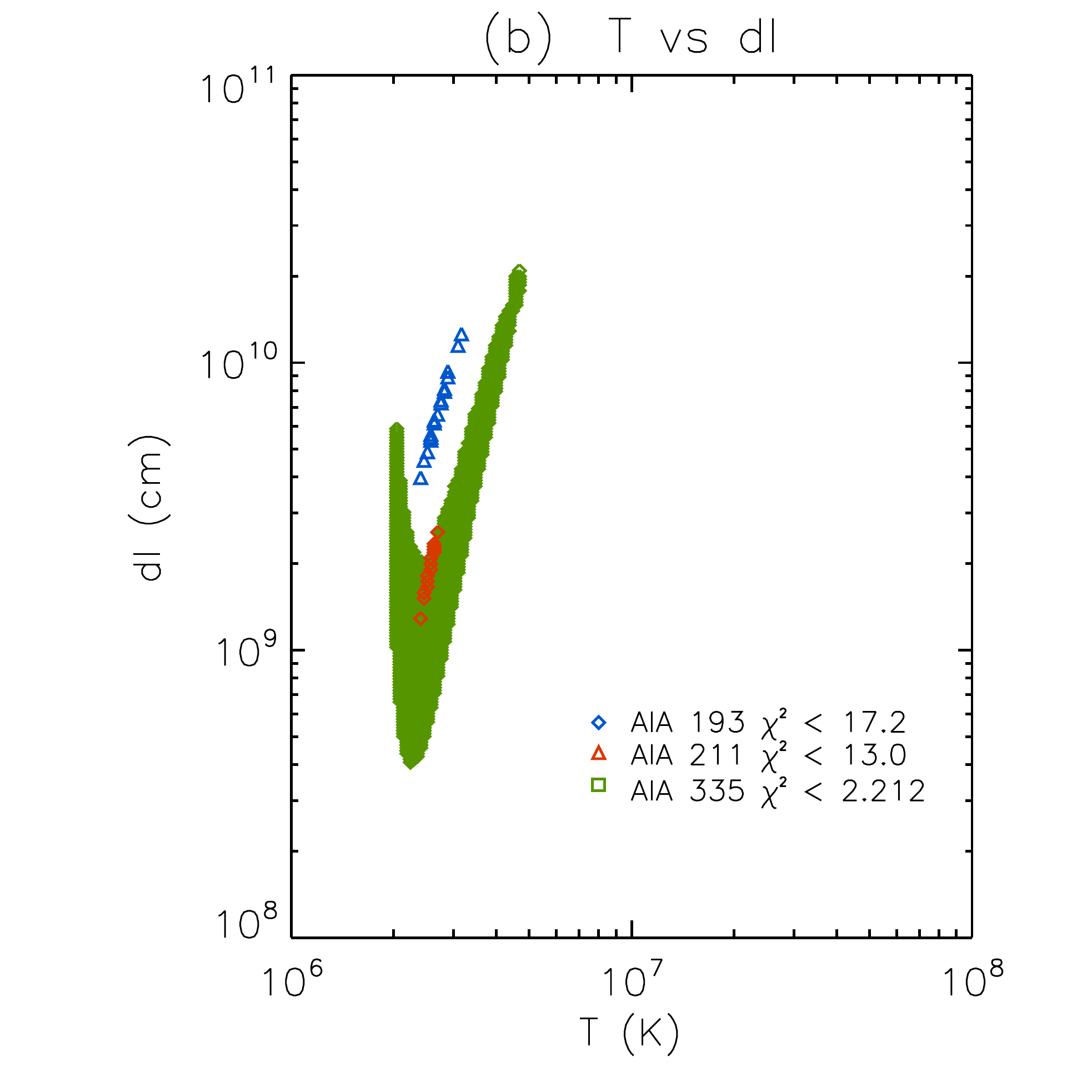}\\
\includegraphics[width=80mm]{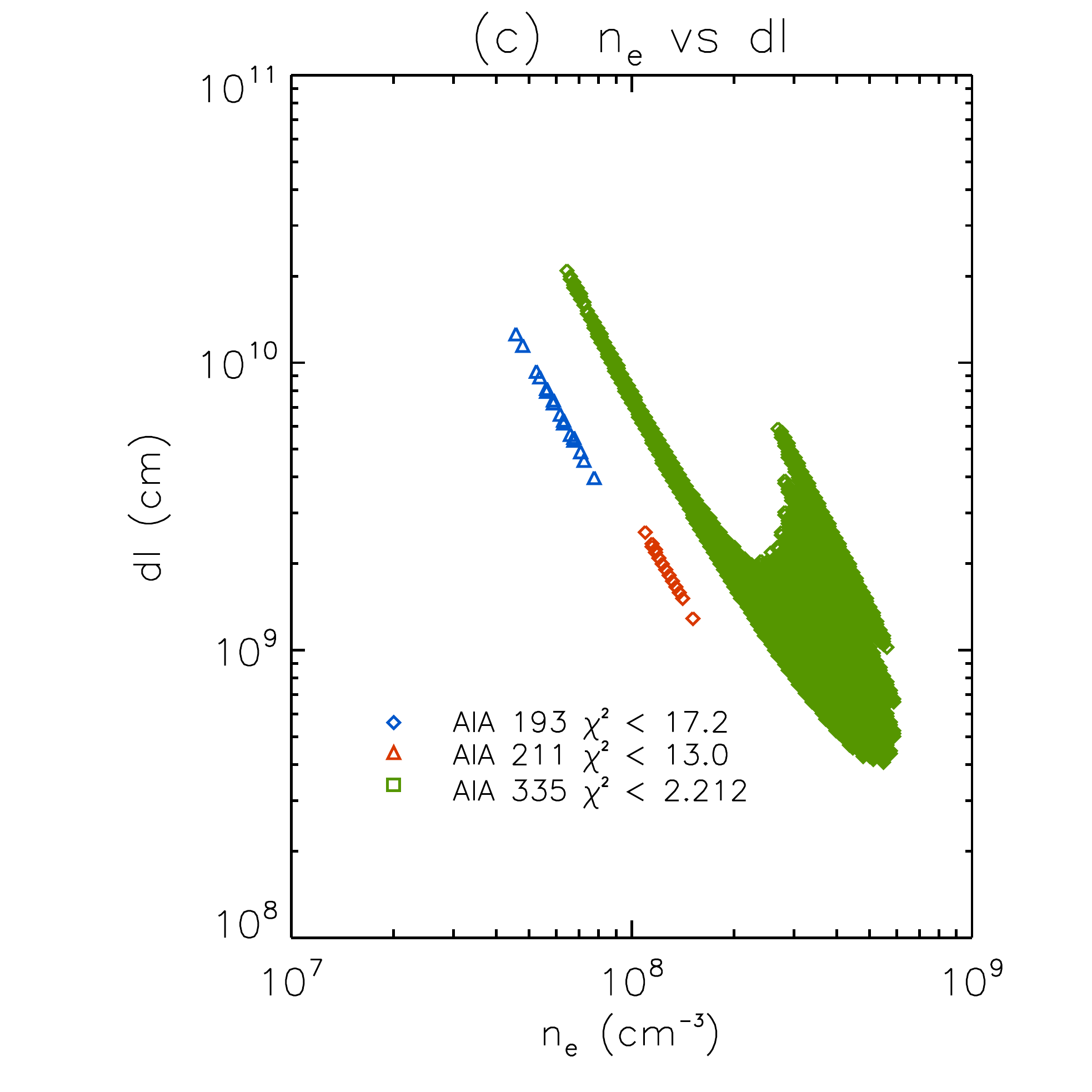}
\caption{Constraints of temperature, density, and line of sight depth 
(a) Temperature vs. Density (b) Temperature vs. Line of sight depth (c) Density vs. Line of sight depth }
\label{fig:shock_cons}
\end{figure}

We show the constraints on T, $n_e$, and $dl$ for each passband in Figure~\ref{fig:shock_cons} and Table~1. 
There is no constraint that satisfies all three passband observations. 
However, there is a clear indication that temperatures of
about 2.4 to 2.7 MK are preferred, while the widths of the peaks require a range
of densities of at least $8 \times 10^7$ to $1.2 \times 10^8 \rm ~cm^{-3}$. 
The density and line of sight depth depend on each other, in the sense that the density tends to decrease 
with the increase of the line of sight depth to match the peak of the observations. 
 
It is apparent that no single set of parameters fits all three bands.  
The probable reasons are 
1) the parameters changing during the course of the observation -- 
the shock sweeps up material from different heights as it moves through the corona; 
2) the background subtraction method is not adequate; and  
3) the shock is not a simple planar structure seen edge-on, but is curved.  
The increasing path length through the shocked gas as a function of time that results from the spherical shape of the shock will distort the evolution of the brightness curves, and we believe that this accounts for most of the discrepancy.

\citet{kozarev2011} found temperatures ranging from about 2.0 to 6.8 MK in two regions behind the shock 
using differential emission measure curves (their Figure 3), with most of the emission at the lower temperatures.  
Recent work has modified the response of the AIA 94 \AA\ band \citep{delzanna2015}, which might remove the need for
emission near 6.8 MK.  
Since \citet{kozarev2011} implicitly assumed ionization equilibrium, 
their temperatures would tend to be too low.  
\citet{ma2011} obtained a temperature of 2.8 MK. 
They only considered the ionization time scales to reach the ionization states 
Fe~XII, Fe~XIV, and Fe~XVI that dominate the 193~\AA\, 211~\AA\ and 335~\AA\ bands, respectively, 
and since they did not detect emission from in the 94 \AA\ band, they found no emission near 6.8 MK. 
Ma et al. did not consider the fading of the 335 \AA\ band, which requires a higher temperature to 
ionize Fe~XVI to Fe~XVII and above, but the fading is complicated by the geometrical effects mentioned above.  
Thus overall, the temperature ranges we obtain are in reasonable agreement with those of Kozarev et al. and Ma et al.

The density ranges shown in Figure~\ref{fig:shock_cons} span an order of magnitude centered 
around 1.2$\times 10^8~\rm cm^{-3}$.  
That is in good agreement with the density of about $10^8~\rm cm^{-3}$ obtained from the type II radio burst \citep{kozarev2011, ma2011}.   Our estimates of the line-of-sight depth also span a large range, from below $10^9$
to above $10^{10}$ cm.  Based on the shock heights given by Kozarev et al. and Ma et al., the depth increased from
zero when the shock entered the extraction box, to about $4 \times 10^{10}$ cm at our last data point. Our estimates are
probably low because of the spherical geometry of the shock and perhaps a small thickness of the shell of shocked
material. 

It is clear that a detailed model of the spherical shock is needed to fit the observations of this event in detail, but since our purpose in this paper is to present a general approach to the AIA response to time-dependent ionization, we defer that to a future paper.  However, it is worthwhile to point out that the modest temperature we derive supports the estimate of \citet{ma2011}, and that temperature is well below that expected for a 600 $\rm km~s^{-1}$ shock unless
the electrons are much cooler than the protons (see \citealp{ghavamian2013}) or else much of the shock energy goes into compressing the magnetic field (quasi-perpendicular shock). 

\section{Conclusion} \label{sec:conclusion}

We find the set of passband responses of SDO/AIA and Hinode/XRT 
which can be applied to the rapidly evolving systems such as a current sheet or a shock.  
We calculate the responses as a function of temperature and characteristic timescale for each passband using 
a time-dependent ionization model that performs fast calculations. 
The two dimensional ratio plot against temperature and $n_et$,  the ratio is almost independent of $n_et$, 
but dependent on temperature for both AIA and XRT. 
The ratio-ratio plots between a pair of passband responses allow us to determine temperature by a combination of temperatures in equilibrium 
and how far the observed plasma is from EI in specific cases. 
We also find that most of the ions reach equilibrium at $n_{e}t\leq$10$^{12}$ cm$^{-3}$s. 
The responses used in this analysis can be found in a site\footnote{\label{NEI_response}https://github.com/jlee2005/NEI-Response}. 

We apply our results to the post-flare arcade of 2012 January 27 and the CME-driven shock of 2010 June 13. 
We find that the temperatures of the flare arcade are similar in the previous work. 
However, the temperature of hotter plasma in NEI is lower than the temperature calculated by assuming EI. 
For the shock, we find that temperatures of
about 2.4 to 2.7 MK are preferred, while the widths of the peaks require a range
of densities of at least $8 \times 10^7$ to $1.2 \times 10^8 \rm ~cm^{-3}$. 
The temperature ranges we obtain are in reasonable agreement with previous works. 
However, a detailed model of the spherical shock is needed to fit the observations. 

We apply our model to the post-flare arcade and the CME-driven shocked plasma with two different methods. 
The method of ratio-ratio plots applied to the post-flare arcade can be used for most of the events 
if the observed DNs have a good signal to noise for each passband. 
However, the method of comparison the observed light curves to the responses in times applied to the shocked plasma 
can be used only for the event that happens for a short time, 
i.e. the observations show an increase and decrease in light curves, 
such as the shocked plasma in this analysis. 

We prepare a robust tool to investigate the physical properties of the plasma in rapidly evolving systems. 
We expect that it could contribute to understanding more quantitatively the evolution of erupting solar events. 

\acknowledgments

CHIANTI is a collaborative project involving George Mason University, 
the University of Michigan (USA) and the University of Cambridge (UK). 
``Hinode is a Japanese mission developed and launched by ISAS/JAXA, with NAOJ as
domestic partner and NASA and STFC (UK) as international partners. It is
operated by these agencies in co-operation with ESA and the NSC (Norway)." 
This work was supported by Basic Science Research Program 
(NRF-2016R1A6A3A11932534) through the National Research
Foundation (NRF) funded by the Ministry of Education of Korea, the Korea
Astronomy and Space Science Institute under the R\&D
program, Development of a Solar Coronagraph on the
International Space Station (Project No. 2017-1-851-00),
supervised by the Ministry of Science and ICT, 
the BK21 plus program through the National Research Foundation (NRF) 
funded by the Ministry of Education of Korea, 
and NRF of Korea Grant funded by the Korean Government 
(NRF-2013M1A3A3A02042232 and NRF-2016R1A2B4013131). 
KR supported by grant 80NSSC18K0732 from NASA to SAO. 
 
\bibliographystyle{aasjournal} \bibliography{nei1}

\begin{deluxetable}{cccc} 
\tabletypesize{\scriptsize}
\tablecaption{Constraints on Temperature, density, and line of sight depth} 
\tablehead{
\colhead{Passband} & \colhead{Temperature (MK)} & \colhead{Density (cm$^{-3}$)} & \colhead{Line of sight depth (cm) } }
\startdata
193 \AA & 2.4 $\sim$ 3.2    & 4.6 $\times$10$^7  \sim$ 7.8 $\times$10$^7$  &  4.0 $\times$10$^9  \sim$ 1.3 $\times$10$^{10}$  \\
211 \AA & 2.4 $\sim$ 2.7    & 1.1 $\times$10$^8  \sim$ 1.5 $\times$10$^8$  &  1.3 $\times$10$^9  \sim$ 2.6$\times$10$^{9}$  \\
335 \AA & 2.0 $\sim$ 4.7    & 6.5 $\times$10$^7  \sim$ 5.9 $\times$10$^7$  &  4.1 $\times$10$^8  \sim$ 2.1 $\times$10$^{10}$  \\
\enddata 
\label{tab:cons}
\end{deluxetable}

\end{document}